\documentclass[a4paper,11pt,times, a4paper, astrosym]{aastex631}
\usepackage[english]{babel}
\usepackage{amsmath}
\usepackage{graphicx}
\usepackage{natbib}

\submitjournal{ApJ}

\shorttitle{Ionization rate from direct sampling of ISN atoms} 

\newcommand{\kms}{~km~s$^{-1}$}

\newcommand{\myvec}[1]{\ensuremath{\boldsymbol{{#1}}}}
\newcommand{\vB}{\ensuremath{\myvec{v}_\text{B}}}

\begin{document}

\title{Determining the ionization rates of interstellar neutral species using direct-sampling observations of their direct and indirect beams}

\correspondingauthor{M. Bzowski}
\email{bzowski@cbk.waw.pl} 

\author[0000-0003-3957-2359]{M. Bzowski}
\affil{Space Research Centre PAS (CBK PAN), Bartycka 18a, 00-716 Warsaw, Poland}

\author[0000-0002-5204-9645]{M.A. Kubiak}
\affil{Space Research Centre PAS (CBK PAN), Bartycka 18a, 00-716 Warsaw, Poland}

\author[0000-0002-2745-6978]{E. M{\"o}bius}
\affil{University of New Hampshire, Durham, NH}

\author[0000-0002-3737-9283]{N.A. Schwadron}
\affil{University of New Hampshire, Durham, NH}

\begin{abstract}
A good understanding of the ionization rates of neutral species in the heliosphere is important for studies of the heliosphere and planetary atmospheres. So far, the intensities of the ionization reactions have been studied based on observations of the contributing phenomena, such as the solar spectral flux in the EUV band and the flux of the solar wind protons, alpha particles, and electrons. The results strongly depend on absolute calibration of these measurements, which, especially for the EUV measurements, is challenging. Here, we propose a novel method of determining the ionization rate of neutral species based on direct sampling of interstellar neutral gas from two locations in space distant to each other. In particular, we suggest performing observations from the vicinity of Earth's orbit and using ratios of fluxes of ISN He for the direct and indirect orbits of interstellar atoms. We identify the most favorable conditions and observations geometries, suitable for implementation on the forthcoming NASA mission Interstellar Mapping and Acceleration Probe. 
\end{abstract}

\section{Introduction}
\label{sec:intro}
\noindent
Neutral atoms enter the heliosphere from the local interstellar medium and bring information on the physical state of the local interstellar matter. However, inside the heliosphere the interstellar neutral (ISN) population of atoms is partly attenuated by ionization. Therefore, a good understanding of ionization reactions of ISN species inside the heliosphere is needed for interpretation of measurements of interstellar and energetic neutral atoms observed by space probes such as Interstellar Boundary Explorer \citep[IBEX; ][]{mccomas_etal:09a} and the planned Interstellar Mapping and Acceleration Probe  \citep[IMAP; ][]{mccomas_etal:18b}. The ionizing reactions include charge exchange (c-x) with solar wind protons and alpha particles, solar wind electron impact (el-imp), and photoionization by solar EUV photons \citep{rucinski_etal:96a}. Typically, the rates of these reactions are calculated based on measurements of relevant physical quantities, like solar wind particle fluxes or solar radiation spectrum \citep[for review, see][]{bzowski_etal:13a}. Clearly, this method of determination of the ionization rates relies on absolute calibration of measurements of the contributing factors. 

Absolute calibration of space-borne instruments has always been challenging for the EUV and  plasma measurements. For solar wind plasma observations, an illustration of the related challenges is the complex effort of maintaining a homogeneous time series of solar wind parameters based on in-situ observations close to 1 au performed over several decades by multiple instruments on various spacecraft, collected in the OMNI2 time series \citep{king_papitashvili:05}. For EUV, the challenges are well illustrated by the attempts of homogenization of the Mg IIc/w time series from different spacecraft \citep{snow_etal:19a}, the Lyman-$\alpha$ time series \citep{machol_etal:19a}, the solar spectrum observed by TIMED \citep{woods_deland:21a} and SOHO \citep{wieman_etal:14a}, and first of all by the complex attempt to cross-calibrate various EUV observations in space \citep[see, e.g., ][]{snow_etal:13}. Dealing with these issues usually involves correlating the observed quantities with various solar proxies, like the sunspot number, the F10.7 and F30 radio fluxes, or the Mg II core to wing ratio. The proxies have their own calibration challenges \citep{clette_etal:16a, svalgaard_hudson:10a, tapping_valdes:11a, tapping:13a, shimojo_etal:17a}. Therefore, an ability to verify the absolute calibration of the ionization rate magnitude at least for selected species would be welcome, especially with a disturbing evidence recently presented by \citet{swaczyna_etal:22a} that the ionization rate of ISN He may be biased during the solar maximum by as much as 40\%. 

Such a large bias confirmed for He would be significant for all heliospheric neutral species, because the photoionization rate of He is mostly due to the solar 30 nm band, which is also mostly responsible for photoionization of H, O, and N, as well as Ne. This band is also strongly correlated with the solar Lyman-$\alpha$ line. Therefore, on one hand, this hypothetical miscalibration would likely turn out responsible for a bias in the understanding of the distribution of ISN H in the heliosphere and in photochemical processes in planetary atmospheres. 

In this paper, we propose an alternative method to estimate the ionization rates of ISN species. We suggest to use observations of ISN He at 1 au obtained from direct-sampling experiments, such as IMAP-Lo on the planned IMAP mission. This method may be applied in a foreseeable future at least for ISN He. In particular, it should enable an estimate of the bias in the ionization rate of ISN He, hypothesized by \citet{swaczyna_etal:22a}, is real. Since the ionization rate of He is mostly due to EUV photons, this should be enough to verify the existence of a large bias in the absolute calibration of solar EUV measurements. 

We start from a brief discussion of the spatial variation of the ionization rate of ISN He in the heliosphere and formulation of simplified operational models (Section \ref{sec:ionRateDesc}). Then, we discuss orbits of ISN atoms and the relation between the ionization rate and the orbital parameters of individual atoms in the direct and indirect orbits. We derive a relation between ISN fluxes observed on two different DOYs on one hand and the ionization rate of the ISN gas on the other hand. We discuss favorable conditions for inferring the ionization rate from such measurements and point out the potential of using observations of the direct and indirect orbits. We also suggest additional prerequisites for the choice of suitable beam pairs for a successful determination of the ionization rate. (Section \ref{sec:baseline}). We discuss variations of the expected ratios of the fluxes for selected beam pairs due to hypothetical bias in the ionization rate and to electron impact with its radial modulation. Subsequently, we verify the ideas developed based on the simple cold-model approximation by simulations performed using a state of the art hot model, as well as implementation aspects on a forthcoming space mission IMAP (Section \ref{sec:implement}). We discuss the results in Section \ref{sec:discussion} and summarize the findings in Section \ref{sec:conclusions}. A detailed discussion of the effect of the uncertainties of the inflow parameters of the ISN gas is presented in the Appendix.

\section{Ionization rate of interstellar neutrals in the heliosphere}
\label{sec:ionRateDesc}
\noindent
Ionization of the ISN gas inside the heliosphere has been discussed by several authors \citep{rucinski_etal:96a,bzowski_etal:13a, bzowski_etal:13b, bochsler_etal:14a, sokol_etal:19a,sokol_etal:20a} and here we only point out the most salient aspects, focusing on the case of ISN He.
\begin{figure}
\centering
\includegraphics[width=0.4\textwidth]{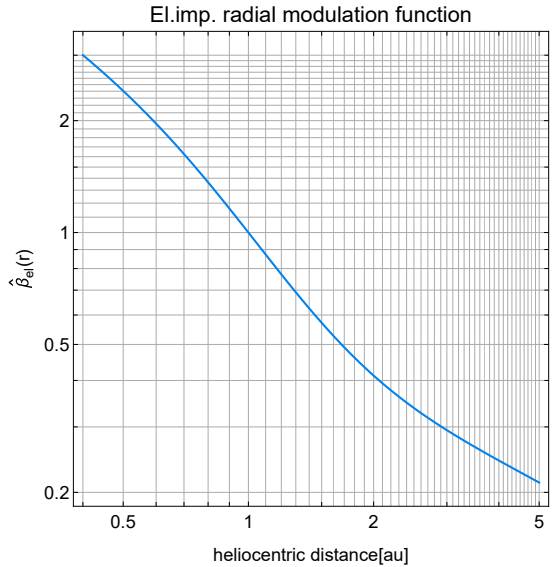}
\caption{The radial modulation function $\hat{\beta}_\text{el}(r)$ of the electron-impact ionization rate, defined in Equation \ref{eq:betaTilde}, as a function of distance from the Sun.}
\label{fig:elImpRadProfile}
\end{figure}

The ionization processes include photoionization by solar EUV photons, solar wind el-imp, and c-x with solar wind protons and alpha particles. All of them vary with time on time scales from minutes to solar cycle period and longer. Additionally, all of them vary with heliolatitude. Concerning the distance from the Sun, photoionization and c-x rates drop with distance squared outside a region where the solar disk cannot be regarded as small (i.e., $\sim 10$ solar radii) and the solar wind is not fully accelerated (several dozens of solar radii). However, the el-imp rate does not follow the same distance law, as demonstrated by \citet{rucinski_fahr:89}. The el-imp rate inside 1 au has not been investigated thoroughly but it is estimated that it drops more rapidly with the solar distance than $r^{-2}$, and that this variation depends on the solar wind regime (fast/slow). 

The dependence on heliolatitude $\phi$ of the c-x and el-imp rates results from the heliolatitudinal structure of the solar wind \citep[see, e.g.,][]{bzowski_etal:13a,tokumaru_etal:21a,porowski_etal:22a}. Also photoionization is, in general, latitudinally-dependent, even though this aspect is much less studied than the solar wind-related variations, and calls for further research \citep{auchere_etal:05c, bzowski:08a, strumik_etal:21b}. Since the trajectories of the atoms discussed in our paper stay close to the ecliptic plane, we will neglect the latitudinal dependence of the ionization rates from now on. 

Effectively, the ionization rate is given by:
\begin{equation}
\beta(r, t) = \left(\beta_{\text{cx},E}(t) + \beta_{\text{ph},E}(t)\right)\left(\frac{r_E}{r}\right)^2 + \beta_{\text{el}}(r, t),
\label{eq:betaDef}
\end{equation}
where $\beta_\text{cx,ph,el}$ correspond to the rates of c-x, photoionization, and el-imp ionization, respectively. The index $E$ marks the rates at a reference distance $r_E$, at which they are measured. Throughout this paper, we adopt $r_E = 1$ au.  

In the case of He, by far the dominant ionization reaction is photoionization. As shown by \citet{sokol_etal:19a}, c-x contributes only $\sim 1$\% to the total rate, which is less than the photoionization uncertainty.  

The topic of el-imp ionization of the ISN gas in the heliosphere was developed by \citet{rucinski_fahr:89} and \citet{rucinski_fahr:91}. They pointed out that because the temperature of solar wind electrons significantly increases towards the Sun, the magnitude of the el-imp rate does not follow the $1/r^2$ relation, which results in an excess rate over that obtained from quadratic scaling of the rate observed at 1 au. \citet{bzowski:08a} and \citet{bzowski_etal:13a} developed a phenomenological formula for this excess rate for ISN H based on solar wind observations from Helios and Ulysses. \citet{bzowski_etal:13b} adapted this derivation for the case of ISN He  (see Equation A.9 with the coefficients in Table A.4 in their paper). The contribution of el-imp ionization to the total ionization of He at 1 au is on the order of 10--20\% \citep{sokol_etal:19a}. While the adopted model of the el-imp ionization rate certainly requires actualization, we adopt it in this paper to illustrate the role of el-imp ionization for the proposed method of determination of the total ionization rate of ISN He. 

The el-imp rate $\beta_\text{el}(r)$ at a distance $r$ is given by a product of the el-imp ionization rate at $r_E$ denoted as $\beta_{\text{el},E}$  and a function $\hat{\beta}_\text{el}(r)$ describing the departure of the radial profile of the el-imp rate from $(r_E/r)^2$:
\begin{equation}
\beta_\text{el}(r) = \beta_{\text{el},E}\, \hat{\beta}_\text{el}(r) \left(\frac{r_E}{r}\right)^2.
\label{eq:betaTilde}
\end{equation}
We will refer to $\hat{\beta}_\text{el}(r)$ as the radial modulation function of the el-imp ionization rate. The el-imp ionization model is linear in the electron density at 1 au $n_e$. For convenience in the further discussion, we express the el-imp ionization rate $\beta_{\text{el},E}$ at $r_E$ for an electron density $n_\text{e}$ as follows: 
\begin{equation}
\beta_{\text{el},E} = n_\text{e}\,\beta_{\text{el},E,1},
\label{eq:figElDensrE}
\end{equation}
where $\beta_{\text{el},E,1}$ is the el-imp ionization rate at $r_E$ from the Sun for a unit electron density. 

The radial modulation function $\hat{\beta}_\text{el}(r)$ used in this paper is presented in Figure \ref{fig:elImpRadProfile} for the relevant solar distance range. As evident from this figure, the el-imp ionization rate at $r = 0.4$ au from the Sun may be as much as three times larger than inferred by scaling the el-imp rate at $r_E = 1$ au by $(r_E/r)^2$. We re-iterate that while this function has been used in the WTPM code since 2008, it certainly needs updating. In our paper, it is used for the purpose of illustration of the ideas being presented.

\section{Baseline idea for determination of the ionization rate from observations of neutral interstellar gas}
\label{sec:baseline}
\noindent
In this section, we present the idea of measuring the ionization rate based on observations of the ISN flux at two different locations in space using the simple cold model of ISN gas. We look for the most favorable combinations of days of the year (DOYs) for the observations performed from Earth's orbit, with the goal of minimizing the observation-based uncertainty of the derived ionization rate. In the assessment of the uncertainties, we include the uncertainty of the measured ISN fluxes and of the inflow parameters of the gas. 

\subsection{Derivation of simplified analytic formulas for the ionization rate and its uncertainties}
\label{sec:analyticDeriv}
\subsubsection{Direct and indirect beams of the ISN gas}
\label{sec:didNdir}
\begin{figure}
\centering
\includegraphics[width=0.5\textwidth]{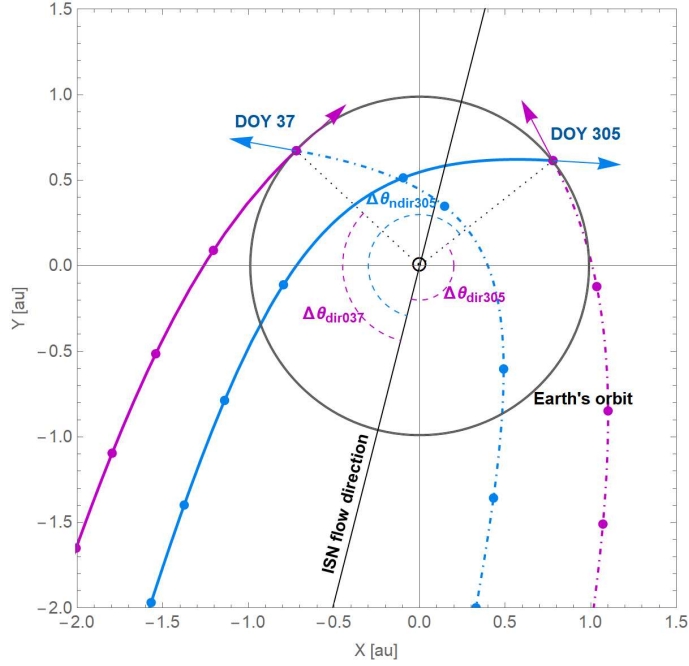}
\caption{The direct and the indirect orbits for two example locations in the Earth's orbit, defined by the DOYs 305 and 37. Solid lines correspond to the orbits preferred for observations on the given DOYs: the indirect orbit for DOY 305, and the direct orbit for DOY 37, as on IBEX. The dash-dotted lines are the complementary orbits for the two selected locations. The indirect orbits are drawn in blue and the direct ones in purple. The arrows mark the directions of the solar-inertial velocity vectors of the atoms. The dots on the trajectories mark time intervals corresponding to the Carrington period. The $\Sun$ symbol marks the Sun, the black Sun-centered circle marks the Earth's orbit, and the dotted lines the radius-vectors of the observer for the selected DOYs. The colored broken arcs mark the angles swept by atoms following the direct and the indirect orbits: $\Delta \theta_{\text{dir305}}$ the angle swept by the atom on the direct orbit, $\Delta \theta_{\text{ndir305}}$ that for the indirect orbit for this DOY, and $\Delta \theta_{\text{dir037}}$ the angle swept by an atom in a direct orbit for DOY 37. }
\label{fig:atomOrbits}
\end{figure}

\noindent
The cold model of the ISN gas flow (\citet{fahr:68, blum_fahr:70a,axford:72}; see the formulation by \citet{thomas:78} and \citet{holzer:77}) assumes that the ISN atoms are collisionless, and far ahead of the Sun (``at infinity'') they flow with identical velocities. Solar gravity bends their trajectories into hyperbolic orbits (see Figure \ref{fig:atomOrbits}) at infinity. 
The ballistic problem has an axial symmetry with the symmetry axis being the atom flow direction. The flow vector \vB{} of ISN atoms and an arbitrary solar radius vector $\myvec{r}$ determine a plane in space. 
Because of the axial symmetry of the ISN flow pattern in the heliosphere, the problem is two-dimensional and the location in the plane defined by $(\myvec{r}, \vB)$ can be parametrized by the length of the radius vector $r = |\myvec{r}|$ and the angle $\theta$ that $\myvec{r}$ makes with the ISN upwind direction given by --\vB, which we will refer to as the offset angle: 
\begin{equation}
\theta =\arccos  (-\vB \cdot \myvec{r})/(|\vB||\myvec{r}|),
\label{eq:thetaDef}
\end{equation} 
where, of course, $0 \leq \theta \leq \pi$.

For any location in space, determined by a radius vector $\myvec{r}$, there are two possible solutions for hyperbolic trajectories with identical velocity vectors at infinity, intersecting at $\myvec{r}$ (see Figure \ref{fig:atomOrbits}). The velocities of atoms in these two orbits at $\myvec{r}$ are denoted as $\myvec{v}_1, \myvec{v}_2$. The orbit with the vector of angular momentum per unit mass $\myvec{r} \times \myvec{v}_i$ pointing above the plane defined by $(\myvec{r}, \vB)$ is referred to as the direct orbit, and the other orbit, with the angular momentum vector pointing below this plane, is referred to as the indirect orbit. 

An example is shown in Figure \ref{fig:atomOrbits}. In this figure, the selected plane is the ecliptic, with the Sun at the crosshairs of the ecliptic reference system. The drawing shows configurations of the direct and indirect orbits in two selected locations on Earth's orbit, denoted by the DOYs when the Earth is in these locations. 
Note that because the two selected points (DOY 37 and DOY 305) are at the opposite sides of the flow direction, the two corresponding offset angles are counted in the opposite directions. The figure is drawn for the actual flow vector of ISN He, adopted from \citet{bzowski_etal:15a}: longitude $75.75\degr$, latitude $-5.16\degr$, and speed $25.764$ \kms. 

The angular momenta per unit mass of the atoms in the direct and the indirect orbits are given by
\begin{equation}
\label{eq:pdirNdirDef}
p_{\text{dir,ndir}} = \frac{1}{2} v_{\text{B}}\left(r \sin \theta \pm \sqrt{r^2 \sin^2 \theta + 4r\, \frac{G M}{v_{\text{B}}^2} (1 - \cos \theta)}\right),
\end{equation} 
where $GM$ is the product of the solar mass and the gravitational constant. The plus sign is for the direct, and the minus sign for the indirect orbit. Regardless of $\theta$, $p_{\text{ndir}} \leq 0$ and $|p_{\text{ndir}}| \leq p_{\text{dir}}$. The density of the ISN gas at $\myvec{r}$ is given by the sum $n_\text{dir} + n_\text{ndir}$, where
\begin{equation}
\label{eq:ndirNdirDef}
n_{\text{dir,ndir}} = \frac{p_{\text{dir,ndir}}^2 w_\text{dir,ndir}}{p_{\text{dir}}^2-p_{\text{ndir}}^2}\, n_\text{LISM};
\end{equation}
$w_\text{dir}, w_\text{ndir}$ are the survival probabilities of the atoms in the direct and indirect orbits against ionization, and $n_\text{LISM}$ is the density of the ISN gas at infinity. The speeds $v$ of the atoms in the direct and indirect orbits at $\myvec{r}$ relative to the Sun are identical because of the energy conservation
even though the directions of the velocities $\myvec{v}_\text{dir}, \myvec{v}_\text{ndir}$ at $\myvec{r}$ are different. 
The fluxes of the direct and indirect beams in the solar-inertial frame at $\myvec{r}$ are given by 
\begin{equation}
\hat{F}_\text{dir, ndir} = v\,n_{\text{dir,ndir}} = n_\text{LISM}\,v\, w_\text{dir,ndir}\,\frac{p_{\text{dir,ndir}}^2 }{p_{\text{dir}}^2-p_{\text{ndir}}^2}.
\label{eq:fluxDef} 
\end{equation}
This equation states that the fluxes related to the direct and indirect beams of ISN atoms at a location $\myvec{r}(r, \theta)$ are proportional to the speed of the atom $v$, the survival probability $w_\text{dir,ndir}$, and a factor 
\begin{equation}
\psi_{\text{dir, ndir}}(\myvec{r}, \myvec{v}_B) = \frac{p_\text{dir, ndir}^2}{p_{\text{dir}}^2-{p_{\text{ndir}}^2}}
\label{eq:psiDef}
\end{equation}
that is fully determined by the velocity vector of the ISN gas at infinity and the observer locus $(r, \theta)$. Hence, the survival probability of the selected beam can be obtained from the formula:
\begin{equation}
  w_\text{dir,ndir} = \frac{\hat{F}_\text{dir,ndir}}{n_\text{LISM}\, v\, \psi_\text{dir,ndir}},
\label{eq:surProFromFlux}
\end{equation}
which depends on the known velocity vector of the atom at infinity and the measured absolute flux of ISN atoms.

\subsubsection{Derivation of the ionization rate based on observations from two vantage points}
\label{sec:betaFm2DOYS}
\noindent
Thus, from the ratio of fluxes $\hat{F}_1, \hat{F}_2$ in different locations 1,2, one can calculate the ratio of survival probabilities $w_1/w_2$:
\begin{equation}
\frac{w_1}{w_2} = \frac{\hat{F}_1}{\hat{F}_2}\,\frac{v_2}{v_1}\, \frac{\psi_2}{\psi_1},
\label{eq:surProRatio}
\end{equation} 
and from this ratio retrieve the ionization rate. It is evident from this formula, that basically, the beams in the positions 1,2 can be selected arbitrarily: both may be either direct or indirect, or one of them may be direct and the other one indirect. It is also possible to stay in one location and use the direct and indirect beam. In practice, however, the choice of the beams and positions is determined by limitations resulting from observation capabilities of the instrument and other factors that will be discussed further on in this paper.

The discussion above has been carried out for an idealized situation of a stationary detector. In practice, measurements are carried out by a detector moving around the Sun with a velocity $\myvec{v}_\text{obs}$. As a result, the observed fluxes are modified. While the measured flux is $F(r, \theta)$, the true flux $\hat{F}$ at $(r, \theta)$ is given by 
\begin{equation}
\hat{F}(r, \theta) = F(r, \theta)\frac{v(r)}{v_{\text{rel}}({r, \theta})},
\label{eq:CGFlux}
\end{equation}
where the velocities of the atoms at $\myvec{r}$ in the detector-inertial frame are given by $\myvec{v}_\text{rel}(\myvec{r}) = \myvec{v}(\myvec{r}) - \myvec{v}_\text{obs}(\myvec{r})$. 
Using this relation and the ratio of ISN fluxes ${\cal{F}}_{1,2}=F_1/F_2$ actually measured at two locations 1,2, the survival probability ratio can be calculated using the formula:
\begin{equation}
  \frac{w_1}{w_2}=
	  {\cal{F}}_{1,2}\,
   \frac{v_{\text{rel},2}}{v_{\text{rel},1}}
	 \frac{\psi_2}{\psi_1}.
\label{eq:surProRatio1}
\end{equation}
The beam speeds in the solar-inertial frame depend solely on the distance from the Sun. 

The theory of survival probabilities for neutral atoms on Keplerian trajectories under time-constant solar conditions was developed by \cite{blum_fahr:70a} and \citet{axford:72}, and for the time-dependent ionization rates by \citet{rucinski_etal:03} (see their Equation 4). Here, we adhere to the convention introduced by \citet{bzowski_etal:13b}. The probability of survival by an atom of its travel from infinity to $\myvec{r}$, i.e., the attenuation factor $w$ for the flux of atoms in a given atom beam observed at $\myvec{r}$, is given by 
\begin{equation}
w(\myvec{r}) = \exp[\cal{E}(\myvec{r})],
\label{eq:wionDef}
\end{equation}
where $\cal{E}(\myvec{r})$ is the exposure to ionization losses, defined as
\begin{equation}
{\cal{E}}(\myvec{r}) = -\int\limits_{t_{\text{TS}}}^{t_{0}} \beta\left(\myvec{r}'\left(t\right),t\right) d t.
\label{eq:exposureDef}
\end{equation}
In this equation, $\beta\left(\myvec{r}'\left(t\right),t\right)$ is the instantaneous total ionization rate at time $t$ at the location $\myvec{r}'\left(t\right)$ on the trajectory. The trajectory starts somewhere far away from the Sun at a time $t_{\text{TS}}$ and ends at a time $t_0$ at the detector located at $\myvec{r}$; $t_{\text{TS}} < t_0$ and $\myvec{r}'(t_0) = \myvec{r}$. The instantaneous ionization rate $\beta(\myvec{r}', t)$ varies because of the changes in the heliocentric distance $r'\left(t\right)$ and heliolatitude $\phi'(t)$ (the spatial effect) and of the evolution in time of the ionizing factors (the time effect). 

For the ionization rate decreasing with solar distance squared and independent of time, $\beta(\myvec{r})=(r_E/|\myvec{r}|)^2\beta_0$, the exposure is proportional to the angle swept by the atom from infinity to $\myvec{r}$ and the ionization rate at 1 au \citep{lee_etal:12a}. Then, it can be calculated analytically, and Equation \ref{eq:exposureDef} simplifies as follows: 
\begin{equation}
		{\cal{E}}_\text{dir,ndir}=-\beta_0 r_{\text{E}}^2 \frac{\Delta\theta_\text{dir,ndir}}{p_\text{dir,ndir}},
		\label{eq:epsAnal}
\end{equation}
where $\beta_0$ is the ionization rate at $r_E = 1$ au, $p_\text{dir,ndir}$ is the angular momentum, and $\Delta \theta_\text{dir,ndir}$ is the angle swept by the atom from infinity to $\myvec{r}$. For the direct and indirect orbits, 
\begin{eqnarray}
  \Delta \theta_\text{dir}   &=& \theta \nonumber \\
  \Delta \theta_\text{ndir} &=& 2 \pi - \theta.
\label{eq:angleSweptDef}
\end{eqnarray}

Hence, the difference between the exposures at 1,2 is obtained as:
\begin{equation}
  {\cal{E}}_1 - {\cal{E}}_2 = -\beta_0\, r_E^2 \left(\frac{\Delta \theta_1}{p_1} - \frac{\Delta \theta_2}{p_2}\right).
	\label{eq:exposureDiff2}
\end{equation}
Combining this equation with Equations \ref{eq:surProRatio1} and \ref{eq:wionDef} we notice that $\ln( w_1/w_2) = {\cal{E}}_1-{\cal{E}}_2$ and obtain an estimate of the ionization rate at 1 au
\begin{equation}
  \beta_0(v_B, \lambda_B, \phi_B, {\cal{F}}_{1,2}) = \frac{
	  \ln \left[
		   {\cal{F}}_{1,2} \frac{\displaystyle v_{\text{rel},2}}{\displaystyle v_{\text{rel},1}} \frac{\displaystyle \psi_2}{\displaystyle \psi_1}
		    \right]
  }
  { r_E^2\left(\frac{\displaystyle \Delta \theta_2}{\displaystyle p_2} - \frac{\displaystyle \Delta \theta_1}{\displaystyle p_1} \right)}.
\label{eq:beta0Def}
\end{equation}

Equation \ref{eq:beta0Def} informs that, with the inflow parameters of the ISN gas established with reasonable accuracy, it is sufficient to measure the flux ratio of selected ISN beams in two locations on Earth's orbit to calculate $\beta_0$. 
This opens interesting possibilities to study various aspects of the ISN gas ionization, especially of the abundant ISN He, with an appropriate selection of the two locations in the Earth's orbit and of the direct or indirect beam in each of them. 
Of course, the right-hand side of Equation \ref{eq:beta0Def} becomes undetermined when one selects the same beams and DOYs because then the denominator disappears.  

The derivation of the uncertainties of the derived ionization rate due to uncertainties in the ISN flow velocity vector and the beam measurement are discussed in detail in the Appendix. 
The conclusions from this analysis are that the uncertainty strongly decreases with increasing difference in the angles swept out along the ISN trajectories of the selected orbit pairs, and is lowest when the selected beam pair is composed of a direct and an indirect orbit.
The contribution from the uncertainty of the ISN flow parameters is insignificant.

\subsection{Selection of the beams and locations in space to minimize the ionization rate uncertainty}
\label{sec:selBestDOYBeams}
\noindent

\subsubsection{Energies of the direct and indirect beams in the spacecraft-inertial reference frame}
\label{sec:beamEnergies}
\noindent
An important aspect of the beam pair choice is the uncertainty in the instrument calibration. 
We mean here the detection efficiency as a function of the energy of atoms impacting the detector. 
While efforts to perform a reliable calibration are made by the instrument teams, there is a remaining instrument-related uncertainty of the observed beam ratio if the selected beams impact the instrument at different energies. 
Therefore, it seems preferable with respect to uncertainty minimization to use beams with identical energies at the detector. 

So far, we have shown that for determining the ionization rate of the ISN gas, it is advantageous to use ratios of the indirect to direct beams. 
Here, we further constrain the DOY pairs suitable for determination of the ionization rate based on the requirement of identical or similar energies of the ISN atoms when impacting the detector. 
We also request the energies to be large enough to exceed the detection threshold for the instrument. 

\begin{figure}
\centering
\includegraphics[width = 0.45\textwidth]{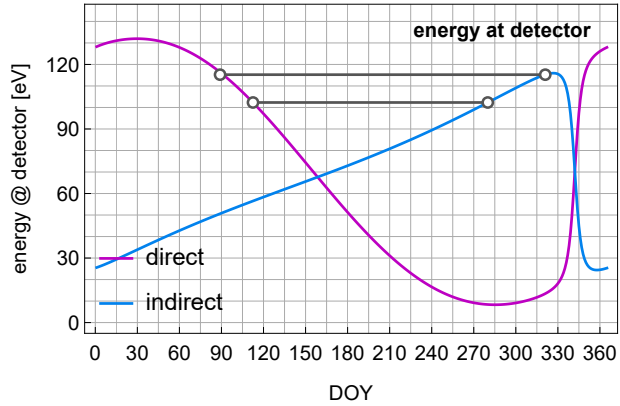}
\caption{Kinetic energy of ISN He atoms from the direct (purple) and indirect beams (blue) relative to the detector traveling with the Earth, shown as functions of DOY. The two horizontal lines connect extreme DOY pairs for the direct and indirect beams for which the atom energy at the detector are identical between the direct and indirect beams (the first and last rows of Table \ref{tab:tabSimu}). The calculations were made using the cold-model approximation.}
\label{fig:plEnergies}
\end{figure}

\citet{sokol_etal:15a} and \citet{galli_etal:15a} demonstrated that the IBEX-Lo sensor has an energy threshold of $\sim 20-30$ eV for detection of ISN He atoms. 
The energies of the direct and indirect beams of ISN He relative to the detector traveling with the Earth for all DOYs over one year are shown in Figure \ref{fig:plEnergies}. 
Clearly, the DOYs for which the direct beam energy is larger than the 20~eV threshold include a range from DOY 1 to~$\sim$ DOY 230, while the time interval for the indirect beam spans DOY 1 to DOY 365. 
Assuming that the IMAP-Lo energy threshold will not be higher than that for IBEX-Lo, we conclude that the energy threshold is not expected to be an important constraint for beam pairs selection. 

Another aspect is a sufficiently large angular distance of the instrument boresight from the Sun. 
IMAP-Lo will be limited to look not closer than 60\degr{} from the Sun. Thus, the impact direction in the spacecraft-inertial frame of the selected indirect beam must be offset from the solar direction by at least this angle.

Based on this consideration we found that the direct-beam DOYs must be selected from the interval $(\sim 89,109)$, and the corresponding DOYs for the indirect beam from the interval $(\sim 289, 321)$. 
The pairing should be done in the increasing order for the direct DOYs and in the decreasing order for the indirect ones. 

\begin{deluxetable*}{CCCcCRCC}
\tablenum{1}
\tablecaption{Corresponding DOYs, elongations, and observer longitudes for the indirect and direct beams. \label{tab:tabSimu}}
\tablewidth{0pt} 
\tablehead{
 \colhead{} & \multicolumn3c{Indirect} & \colhead{} &  \multicolumn3c{Direct} \\
\colhead{Case No} &\colhead{DOY} & \colhead{Elong.} & \colhead{Ecl.longit.} & \colhead{}& \colhead{DOY} & \colhead{Elong.} & \colhead{Ecl.Longit.}} 
\startdata 
 1 & 321 & 64\degr & 55\degr &\vline& 89  & 116\degr & 189\degr\\
 2 & 317 & 64\degr & 51\degr &\vline& 93  & 116\degr & 193\degr \\
 3 & 309 & 60\degr & 43\degr &\vline& 97  & 120\degr & 197\degr\\
 4 & 301 & 60\degr & 35\degr &\vline& 101 & 120\degr & 201\degr\\
 5 & 297 & 60\degr & 31\degr &\vline& 105 & 124\degr & 205\degr\\
 6 & 289 & 60\degr & 23\degr &\vline& 109 & 124\degr & 208\degr\\
\enddata
\tablecomments{The elongation angles are discussed in Section \ref{sec:DOYPairSelect}.}
\end{deluxetable*}
\begin{deluxetable*}{CCCCC}
\tablenum{2}
\tablecaption{Rates of the helium ionization reactions at 1 au in s$^{-1}$ used in the simulations. \label{tab:tabIonRates}}
\tablewidth{0pt}
\tablehead{\colhead{year} & \colhead{photo} & \colhead{el.imp.} & \colhead{ch-x} & \colhead{total}}
\startdata
2015.000 & 1.3029 \times 10^{-7} & 1.50484 \times 10^{-8} & 3.2272 \times 10^{-9} & 1.48565 \times 10^{-7} \\
2009.000 & 5.6196 \times 10^{-8} & 1.18236 \times 10^{-8} & 2.19511 \times 10^{-9} & 7.02147 \times 10^{-8} \\
\enddata
\end{deluxetable*}

\begin{figure}
\centering
\includegraphics[width=0.5\textwidth]{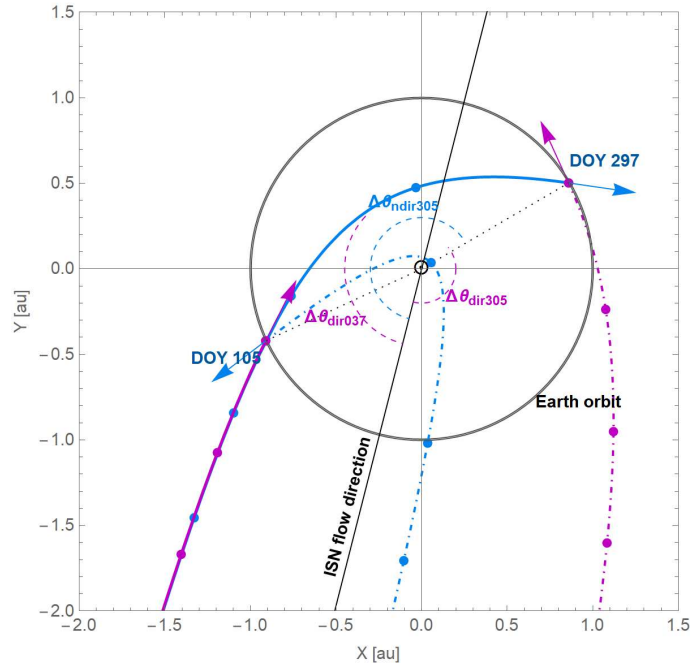}
\caption{Projection of the direct and indirect orbits on the ecliptic plane for a (indirect, direct) DOY pair (case \# 5 in Table \ref{tab:tabSimu}), drawn with the solid lines.
The dash-dotted lines mark the corresponding indirect (blue) and direct (purple) orbits for the two selected DOYs.  
The format of this figure is identical to that of Figure \ref{fig:atomOrbits}. Note that the direct and indirect beams drawn with solid lines overlap before entering inside Earth's orbit. }
\label{fig:plOrbitPairs}
\end{figure}
\subsubsection{Selection of the direct- and indirect beam DOYs and boresight elongations for simulation}
\label{sec:DOYPairSelect}
\noindent
To verify the idea of determining the ionization rate, we will use numerical simulations of the ISN He flux performed using the time-dependent hot-model paradigm \citep{fahr:78, rucinski_etal:03}. 
We will tap into the same simulation base as that used by \citet{sokol_etal:19c}. The available parameter coverage adds further constraints on the selection of suitable DOY pairs in our analysis. 

These simulations are available on a two-dimensional grid for every fourth DOY, i.e., approximately every 4\degr{} in ecliptic longitude, and for every 4\degr{} in the elongation angle\footnote{The elongation angle is the angle between the direction of the spacecraft rotation axis and that of the instrument boresight.}, and also for 90\degr. 
Using Figure \ref{fig:plEnergies} as a guideline, we looked for suitable DOY pairs from the DOY intervals identified in Section \ref{sec:beamEnergies}. 
We found six fitting DOY pairs, listed in the second and fifth columns of Table \ref{tab:tabSimu} for the indirect and direct beams, respectively.  

With the DOY pairs selected, we looked for elongation angles closest to those of the direct and indirect beams shown in Figure 7 of \citet{sokol_etal:19c}. The best-matching elongation angles are listed in columns 3 and 5 of Table \ref{tab:tabSimu}.

\subsection{Differentiation of the survival probabilities of indirect atoms relative to those of direct atoms}
\label{sec:indirSurPro}
\noindent
In this section, we perform a more detailed analysis of the ionization exposures and survival probabilities of the indirect atoms expected inside the observer orbit. We focus on the DOY pairs listed in Table \ref{tab:tabSimu}. 

\begin{figure}
\centering
\includegraphics[width = 0.4\textwidth]{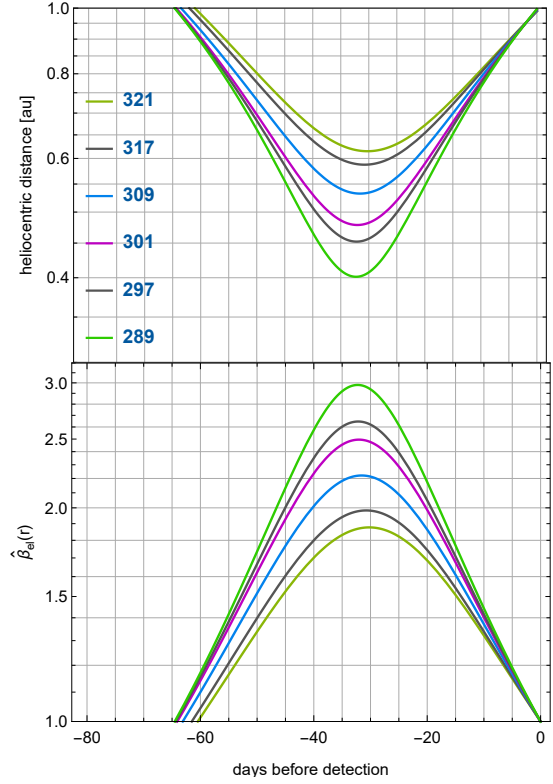}
\caption{Upper panel: distance from the Sun $r(t)$ of indirect atoms inside 1 au as a function of days before detection. The DOYs of detection, listed in the panel, correspond to the indirect DOYs in Table \ref{tab:tabSimu}. Lower panel:  corresponding functions of radial modulation of the electron-impact ionization rate $\hat{\beta}_{\text{el}}(r(t))$, defined in Equation \ref{eq:betaTilde}.}
\label{fig:plnDirOrbit}
\end{figure}

\subsubsection{Ionization exposure on selected indirect orbits }
\label{sec:ioniExpo}
\noindent
The atom orbits for the selected DOY pairs have an interesting property, shown in Figure \ref{fig:plOrbitPairs}: the atoms from the direct beam on a given direct-beam DOY from Table \ref{tab:tabSimu} continue inside the detector orbit and are detected on the indirect-beam DOY from this chosen pair. The travel time inside the detector orbit is shorter than the travel time of the Earth between these two loci in space. As an example, Figure \ref{fig:plOrbitPairs} shows a projection on the ecliptic plane of the direct and indirect orbits for the DOY pair from the 5-th row in Table \ref{tab:tabSimu}. This case was chosen for illustration purposes, but this property holds for all of the entries in this table. If the ionization rates are invariable with time, the ionization exposures $({\cal{E}}_{\text{ndir},i}, {\cal{E}}_{\text{dir},i})$ for the selected orbit pairs $i$ are related as follows:
\begin{equation}
{\cal{E}}_{\text{ndir},i} = {\cal{E}}_{\text{dir},i} + {\cal{E}}_{\text{ndir},i,r < r_E}.
\label{eq:exposureRelation}
\end{equation}
The term ${\cal{E}}_{\text{ndir},i,r < r_E}$ differentiates the survival probabilities of the direct and the indirect atoms. It corresponds to the ionization losses on the portion of the atom orbit inside the detector orbit. 

The upper panel of Figure \ref{fig:plnDirOrbit} presents the heliocentric distances of the indirect atoms for the selected DOYs as a function of time before detection. 
Clearly, these atoms spend about 60---65 days ($\sim 2$ Carrington rotations) inside 1 au, and their perihelia vary from $\sim 0.6$ to $\sim 0.4$ au. 
The latitudes of the perihelia do not exceed 13\degr, which means that the atoms in these orbits stay close to the ecliptic plane. 
For these orbits, the relative excess of the electron ionization rate over the rate obtained from scaling the rates for 1 au using the $(r_E/r)^2$ relation, i.e., the magnitude of the radial modulation function $\hat{\beta}_\text{el}(r)$, is substantial, as illustrated in the lower panel of Figure \ref{fig:plnDirOrbit}. 

\subsubsection{Ionization rate without radial modulation}
\label{sec:IonFlatModulation}
\noindent
In this section, we will discuss ionization of the indirect beam inside 1 au for an ionization rate model with a flat radial modulation function $\hat{\beta}_\text{el} \equiv 1$, meaning that the ionization rate follows $(r_E/r)^2$. 
We focus on the (direct, indirect) beam pairs with identical impact energies, as those selected for analysis in Section \ref{sec:selBestDOYBeams}.

The exposure for the indirect trajectory is given in Equation \ref{eq:exposureRelation}, and since the direct and indirect beams of the selected pairs follow identical trajectories outside $r_E$, the survival probability ratios are identical with the ${\cal{F}}$ ratios for these trajectories. 
 The flux for a given beam can be regarded as a product of the density and speed within this beam. 
We are talking on a flux in the spacecraft-inertial frame. 
We have a beam that first intersects the Earth orbit at the “direct detection” locus and then at the “indirect detection” locus. Physically, this is the same beam. 
By the choice of the detection sites, we know that the speeds related to the beam both in the solar-inertial and spacecraft-inertial frames are identical (in the solar frame, because we are at the close-to-circular Earth orbit, and in the spacecraft-inertial frame due to the selection of the locations in question). 
Consequently, what makes the fluxes different is the local density. 
The reference density is that at the direct-beam location. 
The density at the indirect-beam location is reduced by a factor due to ionization losses. 
Consequently, we have the following relations
\begin{equation}
{\cal{F}} = 
   \frac{F_\text{dir}}{F_\text{ndir}} = 
	 \frac{w_\text{dir}}{w_\text{ndir}} = 
	  \exp\left[{\cal{E}}_\text{dir} - {\cal{E}}_\text{dir} - {\cal{E}}_{\text{ndir}, r<r_E} \right] =
		\exp\left[- {\cal{E}}_{\text{ndir}, r<r_E}\right].
\end{equation}

Let us imagine now that there is a bias in the ionization rate by a factor of $1+\alpha$ in the ionization rate, with $\alpha \ge -1$, but the radial modulation is unchanged. This range of $\alpha$ covers a bias both upwards and downwards. Then
\begin{equation}
{\cal{F}}^\text{biased} = 
   \exp\left[-{\cal{E}}^\text{biased}_{\text{ndir},r<r_E}\right] =
   \exp\left[ -(1 + \alpha)\,{\cal{E}}_{\text{ndir}, r<r_E} \right],
\end{equation} 
that is, the ratio ${\cal{F}}$ is biased by the following factor:
\begin{equation}
\frac{{\cal{F}}^\text{biased}}{{\cal{F}}} = 
\exp \left[ -(1 + \alpha)\,{\cal{E}}_{\text{ndir}, r<r_E}  + {\cal{E}}_{\text{ndir}, r<r_E}\right] =
\exp\left[-\alpha \, {\cal{E}}_{\text{ndir},r<r_E} \right] =
{\cal{F}}^\alpha.
\label{eq:FratiosBiased}
\end{equation}
In other words, a systematic error in the absolute calibration by  factor $1 + \alpha$ of the ionization rate scales proportionally to the magnitude of the observed flux ratio ${\cal{F}}^\alpha$. 
We will verify this conclusion by numerical simulations using the hot model paradigm in Section \ref{sec:implement}. 
The magnitudes of ${\cal{F}}^\text{biased}/{\cal{F}}$ for the ionization models discussed in Section \ref{sec:absCalib} are presented in the lower panel of Figure \ref{fig:pldirndirRatios}.

\subsubsection{Ionization with radial modulation}
\label{sec:ionRadMod}
\noindent
Exposures to ionization losses inside the detector orbit ${\cal{E}}_{\text{ndir},r<e_E}$ are a sum of the exposures to the photoionization and charge exchange rates, which vary with square of the distance to the Sun, and to the el-imp ionization ${\cal{E}}_{\text{ndir,el},r < r_E}$, which features a radial modulation on top of the $(r_E/r)^2$ variation:
\begin{equation}
{\cal{E}}_{\text{ndir},r<e_E} = {\cal{E}}_{\text{ndir,ph},r<r_E} + {\cal{E}}_{\text{ndir,el},r < r_E}.
\label{eq:exposureInsideRe}
\end{equation}
${\cal{E}}_{\text{ndir,ph},r<e_E}$ combines the exposure to photoionization and the small addition from charge exchange. The exposure to photoionization losses inside 1 au is given by 
\begin{equation}
{\cal{E}}_{\text{ph},r<r_E} = -\int\limits_{t_\text{in}}^{t_\text{out}} \beta_{\text{ph,E}} \left(\frac{r_E}{r} \right)^2\, dt,
\label{eq:phIonInside1au}
\end{equation}
where $\beta_{\text{ph},E}$ is the photoionization rate at $r_E$. For this term, Equation \ref{eq:epsAnal} can be used.

Keeping in mind the formulation of the el-imp ionization rate in Equation \ref{eq:figElDensrE}, the exposures to the el-imp ionization inside the detector orbit can be written as 
\begin{equation}
{\cal{E}}_{\text{ndir,el},r < r_E} = - n_e\,\int\limits_{t_\text{in}}^{t_\text{out}} \beta_{\text{el},E,1} \,\hat{\beta}_{\text{el}}(r(t))\, \left(\frac{r_E}{r(t)}\right)^2\, dt = n_e \,{\cal{E}}_{\text{ndir,el},r < r_E,1}.
\label{eq:exposureInside1au}
\end{equation}
where $t_\text{in}$ is the time of the first crossing of the distance $r_E$ from the Sun, $t_\text{out}$ is the detection time, which corresponds to the emergence of the atom from the inside of $r_E$, and $n_e$ is the electron density at $r_E$. 
${\cal{E}}_{\text{ndir,el},r < r_E,1}$ is the exposure to the el-imp ionization inside $r_E$ for the unit electron density. 
We define the survival probability inside $r_E$ against el-imp ionization with unit electron density as $w_{\text{ndir,el},r<r_E,1}$.

If the el-imp ionization would vary with distance as $(r_E/r)^2$, i.e., if $\hat{\beta}_\text{el}(r)\equiv 1$, then the exposure to el-imp ionization inside the detector orbit would be given by:
\begin{equation}
{\cal{\bar{E}}}_{\text{ndir,el},r < r_E} =  -n_e\,\int_{t_\text{in}}^{t_\text{out}} \beta_{\text{el},E,1} \, \left(\frac{r_E}{r(t)}\right)^{\!2} dt = n_e \,{\cal{\bar{E}}}_{\text{ndir,el},r < r_E,1}.
\end{equation}
We use the bar symbols to denote the quantities obtained for the flat radial modulation function.  Both ${\cal{\bar{E}}}_{\text{ndir},r < r_E}$ and ${\cal{E}}_{\text{ndir},r < r_E}$ are negative because $t_{r_{in}} < 0$. Consequently, based on Equations \ref{eq:wionDef} and \ref{eq:exposureRelation}, the survival probabilities for the indirect and direct beams are related to each other in the following way:
\begin{equation}
w_\text{ndir} = w_\text{dir} \exp\left[ n_e\,{\cal{E}}_{\text{ndir,el},r < r_E,1} + {\cal{E}}_{\text{ph},r<r_E}\right] = 
 w_\text{dir} \left(w_{\text{ndir,el},r<r_E,1}\right)^{n_e}\,\exp\left[{\cal{E}}_{\text{ph},r<r_E} \right]
\label{eq:wNdir}
\end{equation}
and similarly for the case when the excess function $\hat{\beta}_\text{el}(r) \equiv 1$:
\begin{equation}
\bar{w}_\text{ndir} = \bar{w}_\text{dir} \left(\bar{w}_{\text{ndir,el},r<r_E,1}\right)^{n_e}\, \exp\left[{\cal{E}}_{\text{ph},r<r_E} \right].
\label{eq:wNdirBar}
\end{equation}
The terms related to photoionization in Equations \ref{eq:wNdir} and \ref{eq:wNdirBar} are identical. The excess attenuation of the indirect beam due to the radial modulation of the electron-impact rate by function $\hat{\beta}_\text{el}$ is given by the ratio $\bar{w}_{\text{ndir,el},r<r_E}/w_{\text{ndir,el},r<r_E}$. This ratio is also equal to the ratio of the ${\cal{F}}$ ratios of direct to indirect beams ${\cal{\bar{F}}}/{\cal{F}}$:
\begin{equation}
\frac{{\cal{\bar{F}}}}{\cal{F}} = \frac{\bar{w}_{\text{ndir,el},r<r_E}}{w_{\text{ndir,el},r<r_E}} = \left( \frac{\bar{w}_{\text{ndir,el},r<r_E,1}}{w_{\text{ndir,el},r<r_E,1}}\right)^{n_e}
\label{eq:wndirBarwndir}
\end{equation} 
While $w_\text{dir} \neq \bar{w}_\text{dir}$ because of the radial modulation of the el-imp rate existing also outside 1 au, we are only interested in the excess ionization losses inside $r_E$. 
The departure of the ratio defined in Equation \ref{eq:wndirBarwndir} informs on the strength of the radial modulation of the el-mp rate. 
Hence, we focus on this quantity now.

\begin{deluxetable*}{LCCCCCC}[!ht]
\tablenum{3}
\tablecaption{Magnitudes of the survival probability ratios defined in Equation \ref{eq:wndirBarwndir} for the (direct, indirect) beam pairs listed in Table \ref{tab:tabSimu}, for various electron densities at the observer orbit.  
\label{tab:lossExcess}}
\tablewidth{0pt} 
\tablehead{
 \colhead{$n_e$ [cm$^{-3}$]} & \colhead{1} & \colhead{2} & \colhead{3} & \colhead{4} & \colhead{5} & \colhead{6} \\
} 
\startdata 
 1           & 1.01286 & 1.01548 & 1.02215 & 1.03119 & 1.03682 & 1.05089 \\  
 n_{e,2009}  & 1.07511 & 1.09101 & 1.13226 & 1.19018 & 1.22749 & 1.32502 \\
 n_{e,2015}  & 1.09703 & 1.11781 & 1.17212 & 1.24934 & 1.29963 & 1.43308 \\
\enddata
\end{deluxetable*}
We performed numerical calculations of ${\cal{E}}_{\text{ndir,el},r < r_E,1}$ and ${\cal{\bar{E}}}_{\text{ndir,el},r < r_E,1}$ for all of the cases defined in Table \ref{tab:tabSimu} adopting a unit electron density and the radial modulation function for $\hat{\beta}_\text{el}$ shown in Figure \ref{fig:elImpRadProfile}. 
Based on these results, the magnitude of the bias seen in the ratio of the direct and indirect fluxes obtained when neglecting the radial modulation function $\hat{\beta}_{\text{el}}(r)$ is given in Table \ref{tab:lossExcess}. 
The table presents the ratios given by Equation \ref{eq:wndirBarwndir} calculated for the reference value of $n_e = 1$ cm$^{-3}$ in the first row, and for electron densities characteristic for 2009.0 and 2015.0: $n_{e,2009} = 5.67$ cm$^{-3}$ and $n_{e,2015} = 7.25$ cm$^{-3}$ in the second and third rows, respectively. 

Clearly, replacing a realistic radial modulation function $\hat{\beta}_\text{el}$ with the flat function results in a larger survival probability inside $r_E$: $\bar{w}_\text{ndir} > w_\text{ndir}$. 
For a reference electron density $n_e = 1$ cm$^{-3}$, the ratio $\bar{w}_{\text{ndir,el},i}/w_{\text{ndir,el},i}$ differs very little from 1 , up to 5\%, as shown in the first row in Table \ref{tab:lossExcess}. 
However, when more realistic electron densities at 1 au are used, the departure of the ratio $\bar{w}_{\text{ndir},i}/w_{\text{ndir},i}$ from 1 increases exponentially as a function of electron density and attains up 25---40\%, as shown for case 6 in Table \ref{tab:lossExcess} for electron densities characteristic for the years 2009 and 2015. 
Based on Equation \ref{eq:wndirBarwndir}, one can obtain the survival probability ratio for different electron densities $n_e$ as $(\bar{w}_{\text{ndir,el},r<r_E,1}/w_{\text{ndir,el},r<r_E,1})^{n_e}$, where the quantity under the exponent is listed in the first row in Table \ref{tab:lossExcess}. 

We conclude from this exercise that el-imp ionization must be addressed explicitly in the analysis of ISN He observations aimed at determining the ionization rate. 
Radial modulation of the el-imp rate may be an important factor shaping the ${\cal{F}}$ ratios of the (direct, indirect) beam pairs. 
Potentially, a series of such measurements could be used to obtain insight into the radial modulation and consequently into the physical state of the solar wind electrons as a function of distance from the Sun inside 1 au.  

In Section \ref{sec:implement} we investigate if this conclusion, based on the cold-model approximation, holds when the thermal spread of the ISN atom velocities is taken into account.

\section{Implementation on IMAP-Lo}
\label{sec:implement}
\noindent
In the previous sections, we suggested an idea to measure the ionization rate of ISN He based on simple cold-model considerations. 
In this section, we evaluate the feasibility of using the IMAP-Lo instrument onboard the forthcoming IMAP mission to obtain the ionization rates based on this idea. 
To this end, we performed simulations using a state-of-the-art hot model of the ISN He flux as seen by the planned IMAP-Lo instrument. 
Note, however, that the goal of our paper is to demonstrate the idea and feasibility of the ionization rate measurement method, and not to make precise predictions.

IMAP will be launched into an orbit around the Sun--Earth Lagrangian point L1. Consequently, it will operate close to the ecliptic plane at $\sim 0.99$ au from the Sun. Its launch is scheduled for 2025, during the maximum of solar activity. 
The IMAP spacecraft will be spin-stabilized, with the rotation axis maintained within a few degrees from the Sun and adjusted daily. 
The IMAP-Lo instrument will be installed on a platform that enables variation of the angle between the spin axis of the spacecraft and the boresight direction of the instrument between 60\degr{} and 160\degr. 

In the following, we investigate the sensitivity of the ratio of simulated fluxes of the direct and indirect beams of ISN He to the absolute magnitude of the ionization rate for the DOYs when the expected energy of the beams at the IMAP spacecraft are almost identical, as listed in Table \ref{tab:tabSimu}. 

\subsection{Simulations}
\label{sec:simulations}
\noindent
Since the details of the future IMAP orbit and the exact launch date are not known yet, our virtual detector was put on Earth's orbit. We simulated the virtually-observed ISN He fluxes for the ISN He inflow parameters adopted from \citet{bzowski_etal:15a} (flow speed 25.784 \kms, flow (longitude, latitude) = $(75.745\degr, -5.169\degr)$, temperature 7443 K) for ionization rate models based on those presented by \citet{sokol_etal:19a}. 
The rates of the contributing ionization reactions at 1 au are listed in Table \ref{tab:tabIonRates}. For the electron-impact ionization rate, we adopted the radial modulation function from \citet{bzowski_etal:13b}, presented in Figure \ref{fig:elImpRadProfile}. 

The simulations were performed using the nWTPM numerical code \citep{sokol_etal:15b} for the observations at the same epoch 2015.0 for all DOY and elongation angle combinations listed in Table \ref{tab:tabSimu}. 
Along with the DOYs, we list ecliptic longitudes of the observer to facilitate the reader to refer the vantage points to the upwind and downwind directions 
The simulation code returns the flux of ISN He in the spacecraft-inertial reference frame, integrated over the atom speeds and the collimator field of view, as a function of the spacecraft spin angle. 
The available simulations  were performed for the centers of the 6\degr{} spin-angle bins in the full range of spin angles from 0\degr{} to 360\degr{}. 
For further analysis, we selected the bins with the energies above the energy threshold of 20~eV and the fluxes above the flux threshold of 100 atoms cm$^{-2}$ s$^{-1}$, similarly to \citet{sokol_etal:19c}. 

To study the sensitivity of the result to different ionization rates, we used the following ionization rate models:
\begin{enumerate}
  \item Full model, with the photoionization, charge-exchange with solar wind protons and alpha particles, and el-imp ionization reactions.
	\item Model as in (1) but with the el-imp rate excluded (further on referred to as ''the photoionization model'')
	\item Two models with the rate adopted from (2) and increased or decreased by $\alpha = \pm 30$\% (two test cases for ``biased ionization rates'').
\end{enumerate}

Models 2 and 3 are composed almost entirely of photoionization because the charge-exchange rates for He are very small. 
Model (2) allows to study the case when the entire ionization rate scales with the square of solar distance. 
Models (3) allow to compare the results and determine the sensitivity of the proposed method to a large bias in the total ionization rate. 
Comparing model (1) with model (2) allows to investigate the sensitivity of the results to the radial modulation of the el-imp ionization rate. 
The difference in the total ionization rates at 1 au between model (1) and (2) is almost exactly 14\%, as can be verified from Table \ref{tab:tabIonRates}.

\begin{figure}
\centering
\includegraphics[width=0.30\textwidth]{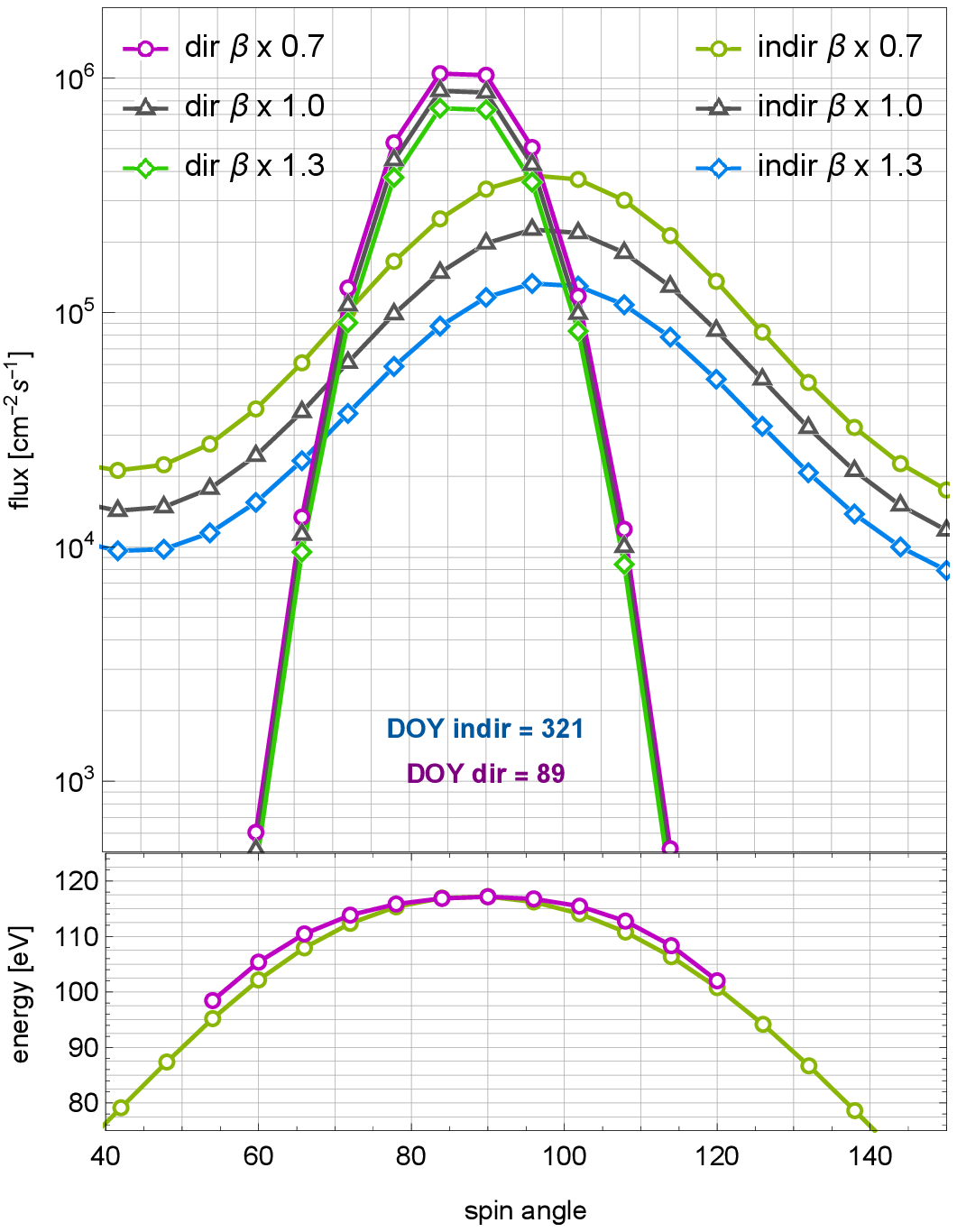}
\includegraphics[width=0.30\textwidth]{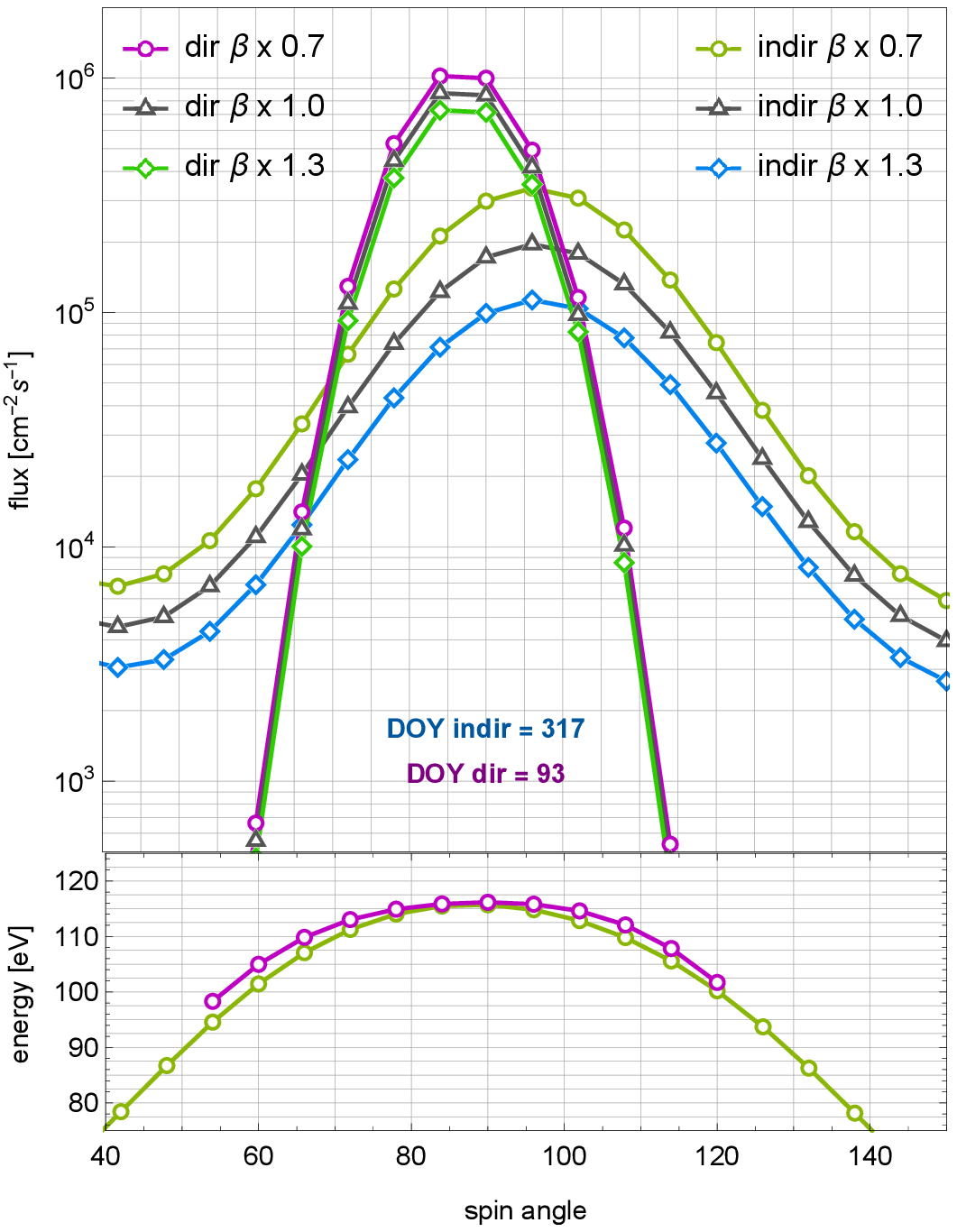}
\includegraphics[width=0.30\textwidth]{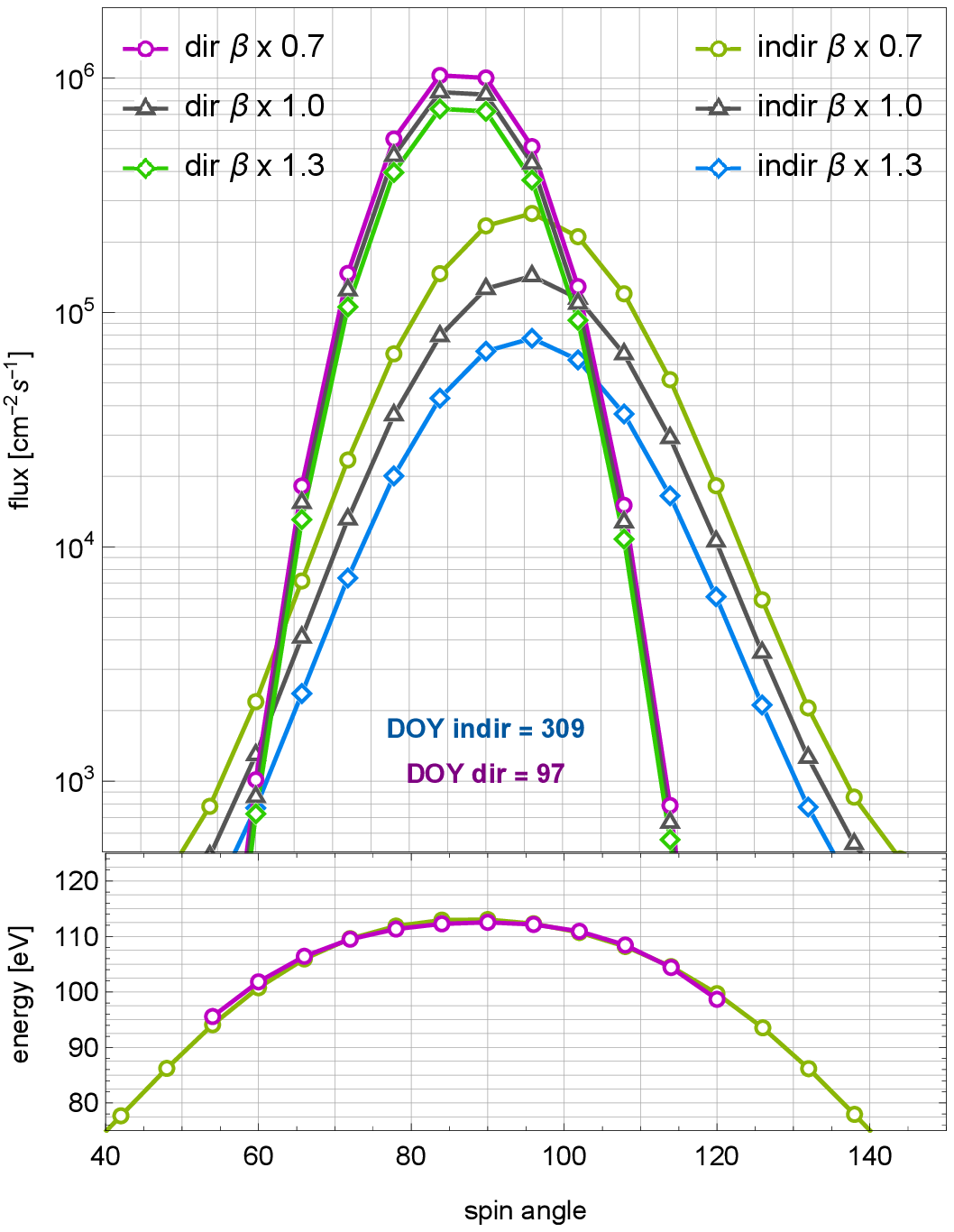}

\includegraphics[width=0.30\textwidth]{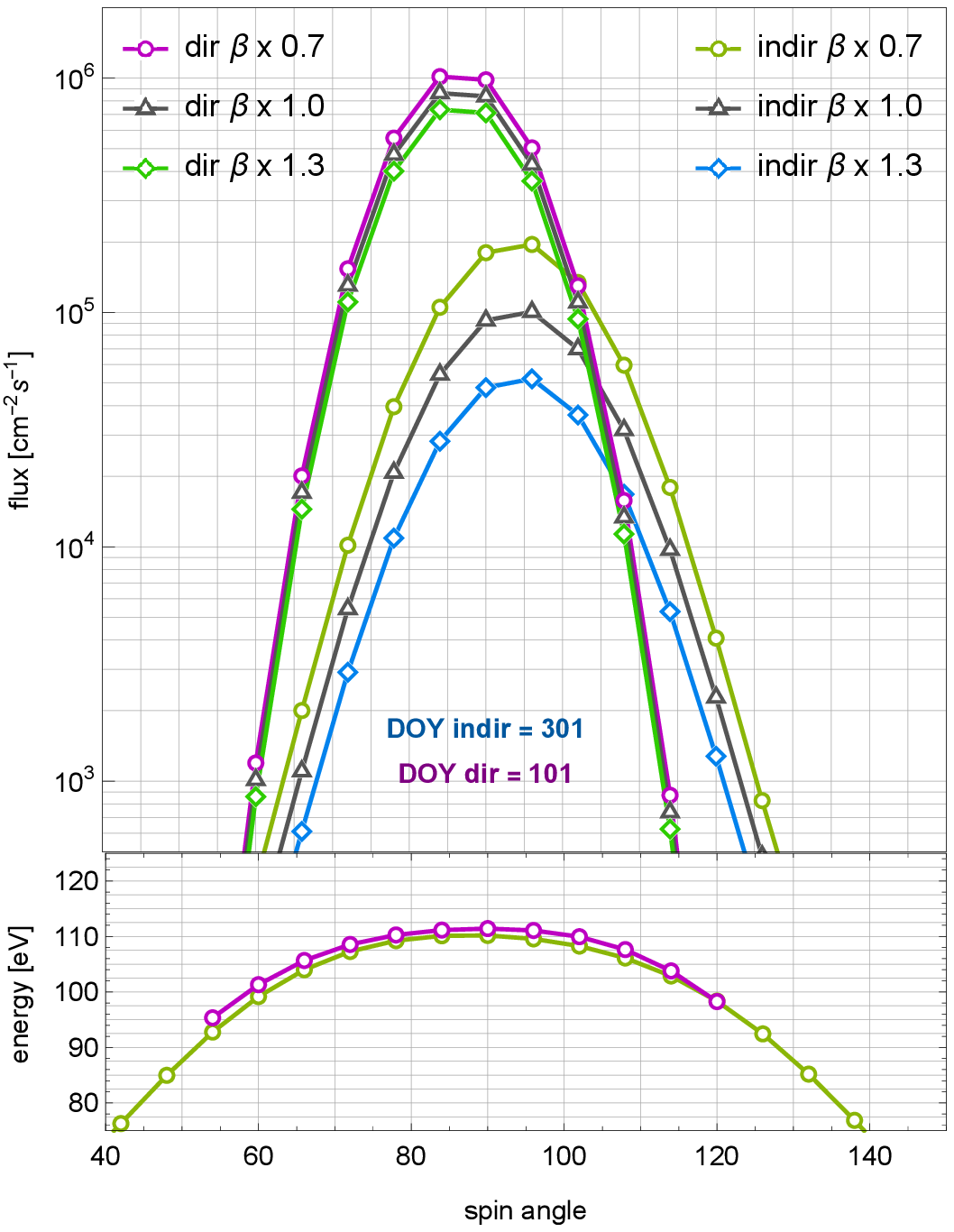}
\includegraphics[width=0.30\textwidth]{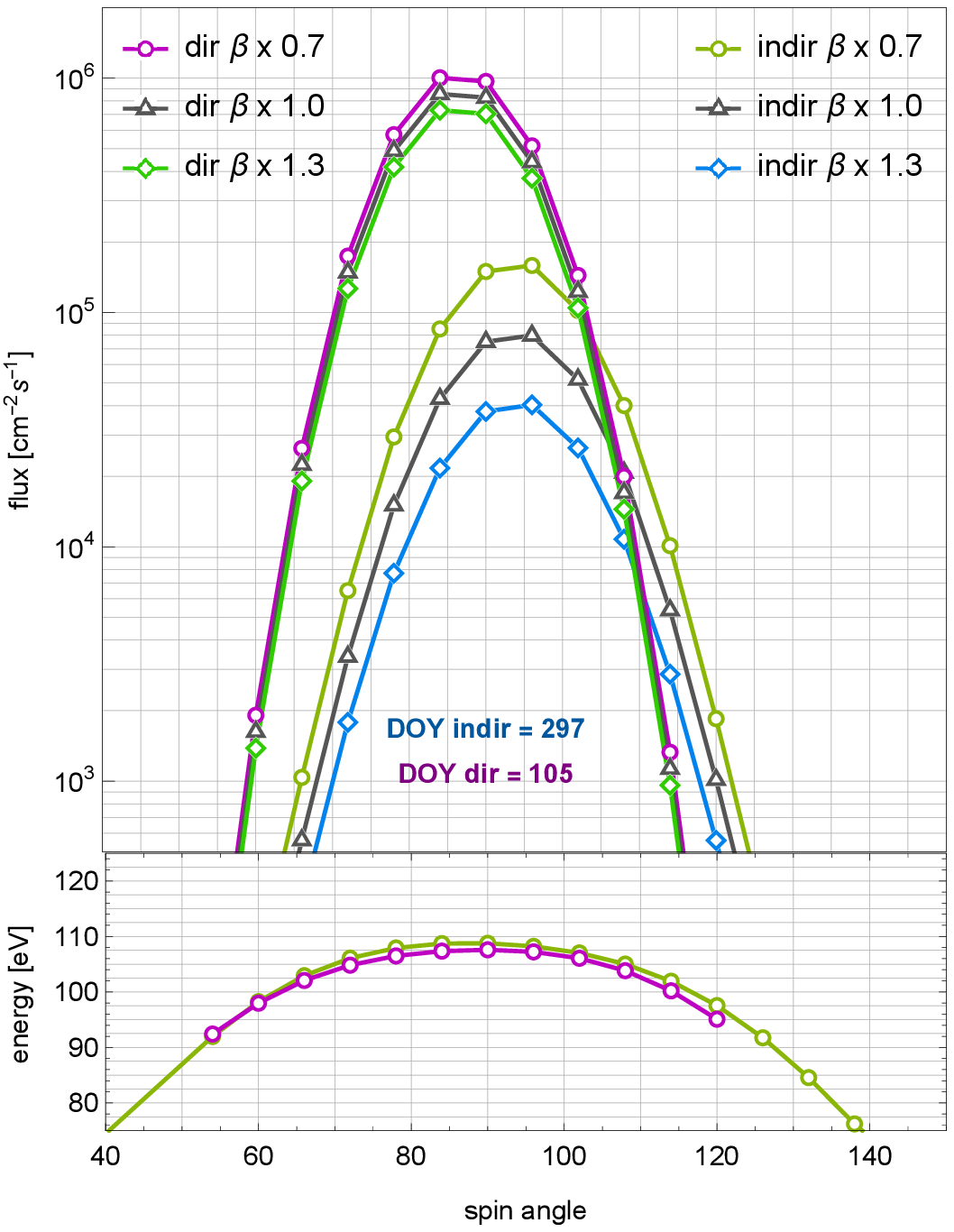}
\includegraphics[width=0.30\textwidth]{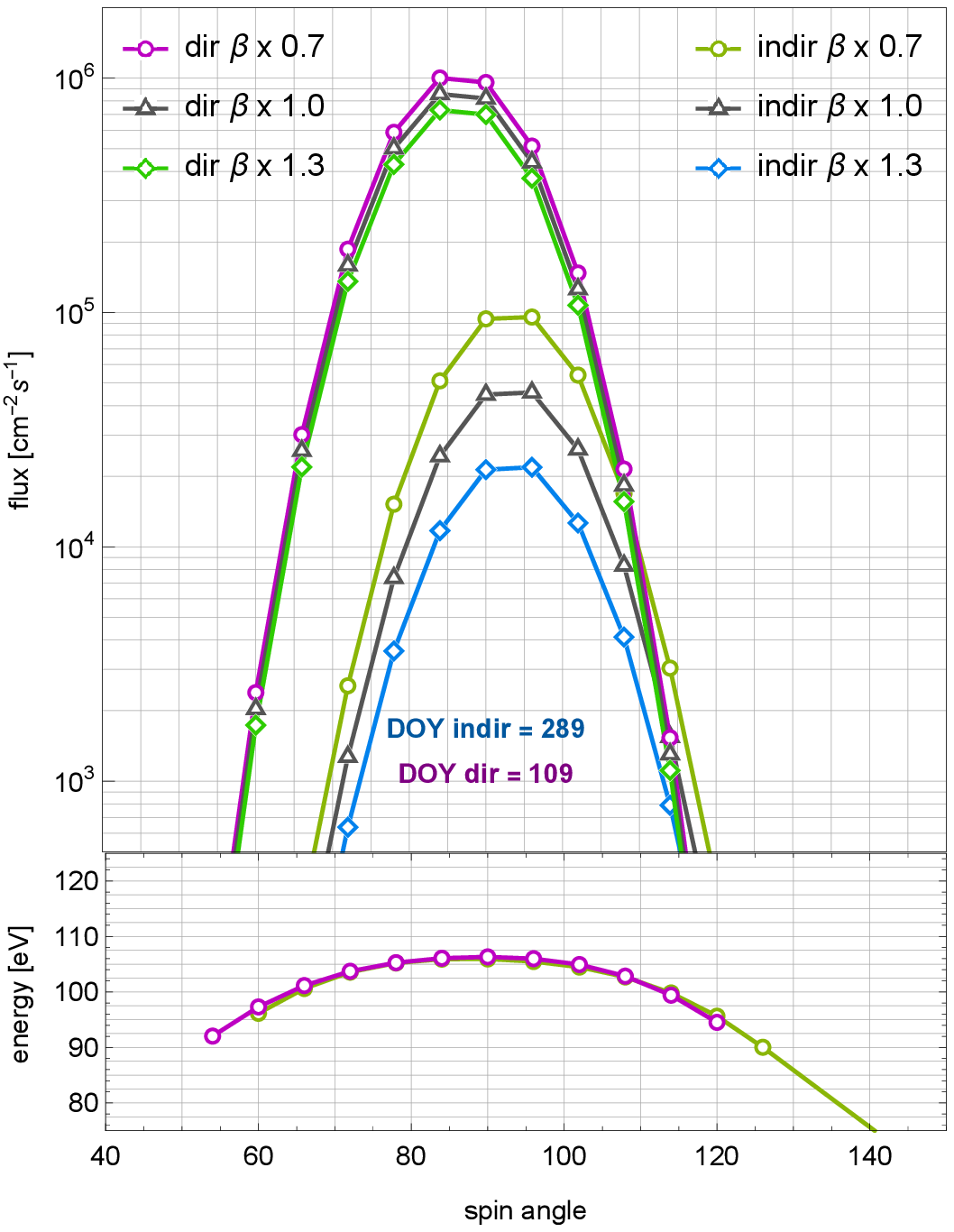}
\caption{Fluxes and energies of the corresponding indirect and direct beams in the spacecraft-inertial reference frame as a function of the IMAP spin angle, for the cases listed in Table \ref{tab:tabSimu}. The upper sub-panels show the absolute fluxes for the corresponding indirect and direct beams, for three different ionization rates: that obtained for photoionization of ISN He from \citet{sokol_etal:19a}, marked as ``$\beta \times 1.0$'', and the rates increased or reduced by 30\% (``$\beta \times 1.3$'' and ``$\beta \times 0.7$'', respectively). The DOY numbers for the indirect and direct beam pairs are listed in the upper subpanels. The lower sub-panels show the corresponding energies of the direct and indirect beams, with the purple color used for the direct beams, and green for the indirect beams. Note that the spin angle range adopted for flux integration is 72\degr{} to 102\degr, i.e., it is much more narrow than that shown in the figure.}
\label{fig:plndirDirFluxes}
\end{figure}

The simulations for the ionization models (2) and (3) are presented in Figure \ref{fig:plndirDirFluxes}. 
They were performed for the conditions corresponding to the high rates characteristic for the solar activity maximum, i.e., for the epoch 2015.0. 
In addition, we repeated some of them for a much more quiet-Sun conditions, characteristic of 2009. The objective was to verify that the suggested method works fine for the ionization rates spanning the entire solar activity cycle. 

Additional simulations were performed for the Warm Breeze. The Warm Breeze, discovered in the IBEX-Lo data by \citet{bzowski_etal:12a} and investigated in greater detail by \citet{kubiak_etal:14a, kubiak_etal:16a, bzowski_etal:17a,bzowski_etal:19a}, and \citet{fraternale_etal:21a}, is the secondary population of ISN He, created in the outer heliosheath. 
It was also discussed in the context of future IMAP-Lo observations by \citet{sokol_etal:19c}.
It has a slower flow speed at the heliopause than that of the primary population, and thus a different energy at 1 au. 
Its abundance at the heliopause is relatively low, about 5\% of that of ISH He, but the signal is well visible in IBEX-Lo data and expected to be visible in the IMAP-Lo observations. 
Therefore, we performed simulations of the expected Warm Breeze flux using a Maxwell-Boltzmann approximation for its distribution function at the heliopause with the canonical set of inflow parameters suggested by \citet{kubiak_etal:16a} to verify that the Warm Breeze will not bias the ionization rates obtained from the analysis suggested in our paper. 

\subsection{Verification of the equal-energy prerequisite and spin angle range selection}
\label{sec:energy}
\noindent
We start with a verification of the equal energy prerequisite for the direct and indirect beams at pairs of specific DOYs. 
First of all, we have verified that different ionization rates do not alter the energies observed in individual spin angle bins. 
Therefore, in the lower subpanels in Figure \ref{fig:plndirDirFluxes} we only show the energies for one of the three (direct, indirect) beam pairs as a function of the spacecraft spin angle. 

Analysis of Figure \ref{fig:plndirDirFluxes} shows that the prerequisite of equal energies for the corresponding indirect and direct beams is fulfilled, especially for the spin angle bins close to the peak energies. 
The small differences visible for some of those bins result mostly from non-ideal alignment in the (DOY, elongation) pairs. 
This is because the (DOY, elongation) pairs in the WTPM simulation base are available at a (DOY, elongation) grid with a resolution of 4 DOYs and 4\degr{} in elongation. 
Constructing ideal pairs would require a finer resolution. 
Nevertheless, inspection of Figure \ref{fig:plndirDirFluxes} suggests that it is possible to directly compare and take ratios of individual (direct, indirect) bins with the same-energy condition fulfilled, and that it is thus possible to compare directly the fluxes integrated over selected ranges of spin angle.

\begin{figure}
\centering
\includegraphics[width=0.45\textwidth]{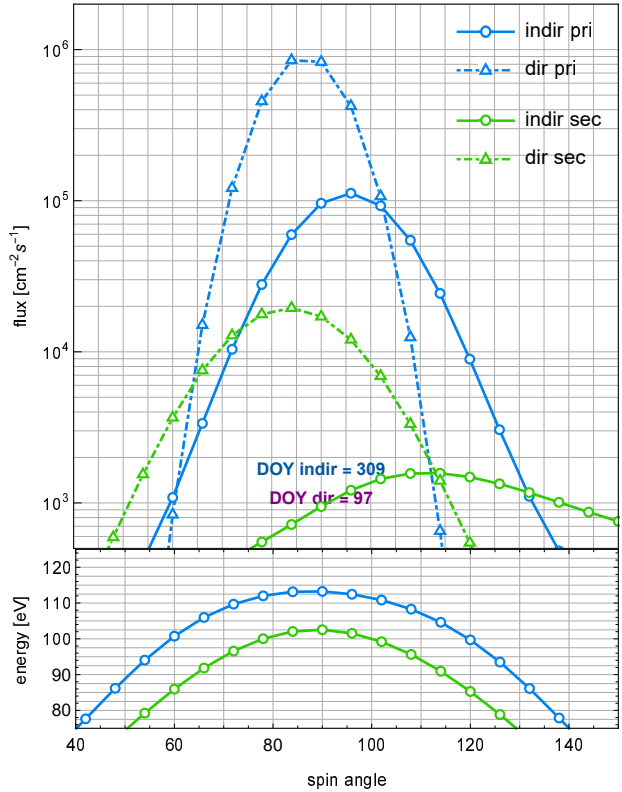}
\caption{Fluxes of the primary and secondary populations of ISN He, direct and indirect beams (upper panel), and the corresponding beam energies (lower panel) for a pair of the (direct, indirect) DOYs (case 3 in Table\ref{tab:tabSimu}). In the lower panel, the blue line represents the energies for the indirect and direct beams of the primary population, and the green line those of the secondary population.}
\label{fig:plPriAndSec}
\end{figure}

The selection of the spin-angle ranges must be done with care, to avoid biasing the result by the Warm Breeze. 
This issue is illustrated in Figure \ref{fig:plPriAndSec}, which presents the direct and indirect beams of the primary and the secondary populations of ISN He for one of the DOY pairs selected for analysis.
The lower panel of this figure shows the corresponding energies. 
Clearly, the fluxes of the secondary population are not negligible compared with those of the primary ISN flow, both for the direct and the indirect beams. 
Moreover, the energies of the secondary population are different to those of the primary population. 
This is understandable, because the inflow parameters of the secondary population are different to those of the primary, so the energies of the two populations at 1 au are different. 
Additionally, the selection of the (DOY, elongation) pairs optimized for the primary population is not optimized for the secondary, so the (direct,indirect) WB beam pair energies in the detector frame are different. 
An unaccounted presence of the Warm Breeze component increases the indirect to direct flux ratio, introducing a bias with the magnitude depending on the choice of the spin angle range used for the flux integration. 
Hence, for the needs of determination of the ionization rate we want to minimize the component of the measured flux from the secondary population.

To that end, we follow the approach first suggested by \citet{bzowski_etal:12a} to restrict the spin angle range so that the primary population dominates. Inspection of Figure \ref{fig:plPriAndSec} from the viewpoint of the direct beam of the primary population suggests that adoption, after \citet{swaczyna_etal:18a} and \citet{swaczyna_etal:22a}, of the spin angle range (72\degr, 102\degr) for the selected DOY pairs is reasonable, and that this range can also be safely adopted for the primary population's indirect beam. This choice limits the influence of the secondary population on one hand and allows to obtain a reasonably good statistics of counts on the other hand. We compared the flux ratios of the direct and indirect beams calculated either with the secondary population included or excluded for the simulations using ionization model (2). We found that integrating the beams over this spin angle range limits the bias of the derived ionization rates toward increased values due to the presence of the secondary population to less than 2\% of the ratio of the direct to indirect fluxes. Extending the adopted spin angle range rapidly increases this bias while enhancing the count statistics very little because of a rapid decrease of the direct beams, as shown in Figures \ref{fig:plndirDirFluxes} and \ref{fig:plPriAndSec}.

\subsection{Validation of the absolute calibration of the ionization rate model}
\label{sec:absCalib}
\noindent
In this section, we study whether the method of indirect beams enables the detection of a hypothetical large systematic bias in the ionization rate model, such as that  hypothesized by \citet{swaczyna_etal:22b} based on their analysis of IBEX-Lo observations. 
When the photoionization rate is calculated based on the observed solar EUV spectra, like those available from TIMED \citep{woods_etal:05a}, the systematic uncertainty of $\sim 10-15$\% of the spectral flux propagates into the ionization rate as a comparable systematic uncertainty. However, \citet{swaczyna_etal:22b} suggested that the actual error in the total ionization rate of ISN He might be more than twice as large. 

\begin{figure}[!ht]
\centering
\includegraphics[width=0.4\textwidth]{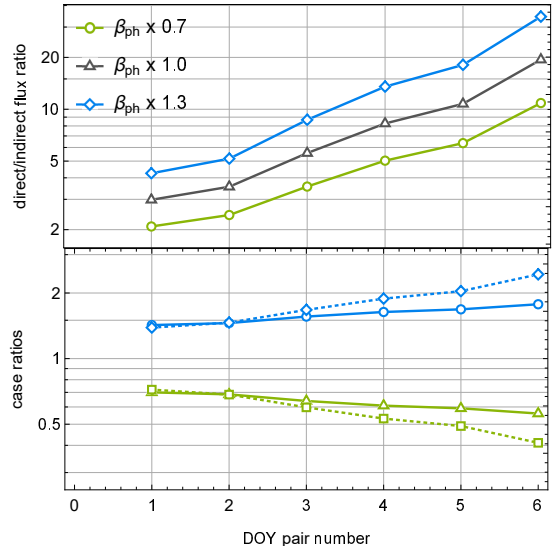}
\caption{Ratios of the spin angle-integrated direct/indirect fluxes (the ${\cal{F}}$ ratios) of ISN He, simulated for the ionization model (2) (``$\beta_\text{ph}$ x 1.0'') and the ${\cal{F}}$ ratios for the ionization models (3), i.e., with the rate increased (``$\beta_\text{ph}$ x 1.3'') and decreased (''$\beta_\text{ph}$ x 0.7'') by 30\%, are shown in the upper panel for the (indirect,direct) DOY pairs listed in Table \ref{tab:tabSimu}. The lower panel  presents the ratios of the ``$\beta_\text{ph}$ x 1.3'' and ``$\beta_\text{ph}$ x 0.7'' lines to the line ``1.0'' (solid lines). The dashed lines in this panel represent the corresponding cold model-derived ratios defined in Equation \ref{eq:FratiosBiased} for $\alpha = \pm 0.3$.}
\label{fig:pldirndirRatios}
\end{figure}

We perform this study by analyzing ratios ${\cal{F}}_i = F_{\text{dir},i}/F_{\text{ndir},i}$ of the simulated spin angle-integrated direct and indirect beams, where $i$ corresponds to the DOY pair number given in the first column of Table \ref{tab:tabSimu}. We avoid using absolute magnitudes because the absolute calibration of the IMAP-Lo instrument may turn out to be not precise enough to draw conclusions based on analysis of absolute fluxes, and furthermore, the estimate of the density of ISN He, to which the observed flux is proportional, has a large uncertainty \citep{mobius_etal:04a}. Therefore, a comparison of the absolute values of the measured fluxes with those simulated may likely be inconclusive. By contrast, the ratios of the observed fluxes are free from these uncertainties and can be directly compared with simulations. In fact, the ionization rate can be obtained from fitting the simulated ${\cal{F}}$ ratios to those observed by varying the absolute ionization rate used in the model. 

To test the sensitivity of our proposed method to the magnitude of the ionization rate, we performed simulations using ionization models (2) and (3) discussed at the end of Section \ref{sec:simulations}. The simulated absolute fluxes for the centers of spin angle bins are presented in Figure \ref{fig:plndirDirFluxes}, where individual panels present the results for the corresponding DOY pairs for the indirect and direct beams, listed in Table \ref{tab:tabSimu}. 

The upper panel of Figure \ref{fig:pldirndirRatios} presents the simulated ${\cal{F}}$ ratios of the direct to indirect beams as a function of the beam pair number listed in the first column in Table \ref{tab:tabSimu}. For larger pair numbers, the indirect beams have increasingly lower perihelion distances and thus larger exposures to ionization losses. Consequently, the magnitude of the ${\cal{F}}$ ratio increases. This is illustrated by the black line in Figure \ref{fig:pldirndirRatios}, which was obtained for the ionization model (2), i.e., assuming that the ionization rate decreases as $(r_E/r)^2$ and the radial modulation function is flat: $\hat{\beta}_\text{el}(r) \equiv 1$.

Now, we introduce an artificial bias to model (2) and take ionization models (3), which are obtained by increasing or decreasing the ionization rate (2) by 30\% while maintaining the same radial profile. The ${\cal{F}}$ ratios for these two cases are shown with the green and blue lines in Figure \ref{fig:pldirndirRatios}. For an increased ionization rate (green line) the ratio is larger because an increased intensity of ionization results in a stronger attenuation of the indirect beam. This effect is also clearly seen in Figure \ref{fig:plndirDirFluxes}. Conversely, for a reduced magnitude of the ionization rate, shown with the blue line in Figure \ref{fig:pldirndirRatios}, the exposure to ionization losses of the indirect beam is reduced, the measured indirect beam is greater, so the ${\cal{F}}$ ratio is reduced.

Clearly, the ratios for the increased and reduced ionization rates strongly differ from those for the nominal photoionization rates, which demonstrates by how much the ${\cal{F}}$ ratio is sensitive to the magnitude of ionization. 
Assuming a flat radial modulation function, the measurement for one pair of (direct, indirect) beams is sufficient to validate the absolute magnitude of the ionization. 

This sensitivity of the ${\cal{F}}$ ratio to the absolute magnitude of the ionization rate is quantified in the lower panel of Figure \ref{fig:pldirndirRatios}, where we show the ratios of the ${\cal{F}}$ quantities obtained for the different ionization rates. 
For the selected spin angle range, the sensitivity to the magnitude of the ionization rate is amplified, i.e., the ${\cal{F}}$ ratios are larger than the ratios of the adopted ionization rates. 

The ratios obtained using the hot model (solid lines in Figure \ref{fig:pldirndirRatios}) are close to the simplified predictions obtained from the cold model (Equation \ref{eq:FratiosBiased}) for the first three pairs of DOYs (shown with broken lines). 
For the subsequent pairs, small systematic discrepancies appear. Since the cold model is only a rough approximation to the actual distribution of the ISN He flux, especially close to the ISN cone region, we recommend to use the hot model and to obtain the ionization rates from numerical fitting of hot-model flux ratios to future observations. 

\subsection{Sensitivity of the ${\cal{F}}$ ratio to electron-impact ionization} 
\label{sec:elIonSensit}
\begin{figure}
\centering
\includegraphics[width=0.45\textwidth]{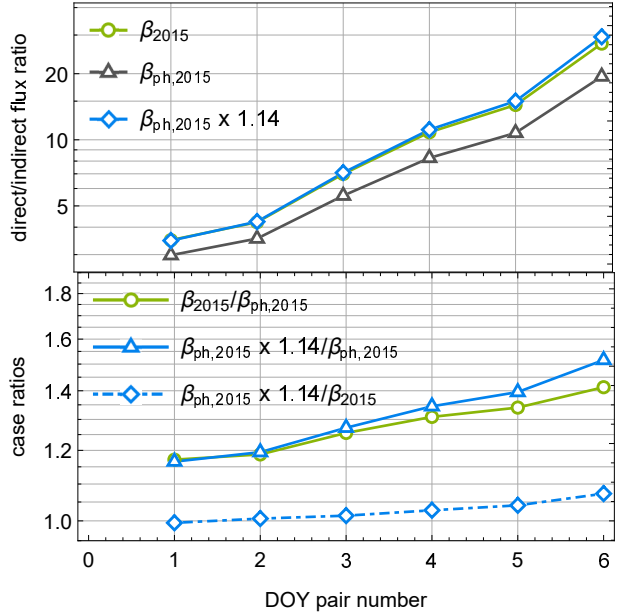}
\caption{Upper panel: the ${\cal{F}}$ ratios for the ionization rates characteristic for the maximum of the solar activity, i.e., for the sum of photoionization and electron-impact rate (``$\beta_{2015}$'') and for the photoionization rate only, with electron-impact ionization neglected, for the solar maximum conditions (``$\beta_{\text{ph},2015}$'').
At 1 au, the electron-impact ionization rate contributes $\sim 14$\% to the photoionization rate, and more than that with a decreasing distance from the Sun. 
The line marked as ``$\beta_{\text{ph},2015}$ x 1.14'' presents the ${\cal{F}}$ ratio simulated for the ionization model where the photoionization rate is increased by 14\% consistently for all distances, and electron-impact ionization is neglected. 
Lower panel: ratios of the respective ${\cal{F}}$ ratios shown in the upper panel. }
\label{fig:plSolMinSolMax}
\end{figure}

\noindent
The proposed method of determining the ionization rate does not differentiate between different ionization reactions and assumes that the radial modulation function of the ionization rate is flat. 
In Section \ref{sec:ionRadMod} we found based on the cold model that the effect of the radial modulation function of the el-imp rate relative to a $r^{-2}$ dependence should be easily detectable in a series of measurements suggested in Table \ref{tab:tabSimu}. 
Potentially, this could offer an opportunity to study the intensity of el-imp at a range of distances inside 1 au and hence to provide information on the physical state of the solar wind electron component. 
Since the cold model is only an approximation, we evaluate here based on a state-of-the-art hot model to what extent the radial modulation of the el-imp rate can be visible in observations planned for IMAP-Lo.

We remind the reader here that the el-imp rate adds $\sim 14$\% to the photoionization and charge exchange rates (for the 2015 conditions) at 1 au. 
Inside 1 au, this contribution is expected to be larger due to the radial modulation. Note that the systematic uncertainty of the photoionization rate of He, reported as $\sim 10 - 15$\%, is close to this value. 

To investigate the detectability of the radial modulation effect in the future IMAP-Lo observations, we take the ${\cal{F}}$ ratios obtained for the full ionization model for 2015, with the radial modulation of the el-imp rate included (Model 1; the green line in Figure \ref{fig:plSolMinSolMax}), and compare it with the ${\cal{F}}$ ratios obtained for the model with the el-imp rate excluded (Model 2; the black line) and with the latter model adjusted for the difference between the full ionization model and that with the el-imp rate omitted by application of Equation \ref{eq:FratiosBiased}. 
For the exponent, we take $\alpha = \beta(r_E)/\beta_\text{ph}(r_E) \simeq 1.14$. 
This latter result is presented with the blue line in the upper panel of Figure \ref{fig:plSolMinSolMax}. 
The difference between the blue and green line is due to the radial modulation of the ionization rate (see the dashed blue line in the lower panel of Figure \ref{fig:plSolMinSolMax}). 

Clearly, adding the el-imp rate to the photoionization rate is immediately visible in the ${\cal{F}}$ ratio: the difference between the black and green line in Figure \ref{fig:plSolMinSolMax} increases from $\sim 15$\% to $\sim 40$\% for the beam pairs 1--6, as shown with the green line in the lower panel of this figure. 
Adoption of a flat radial modulation function results in a different series of ${\cal{F}}$ ratios, as illustrated with the blue line in Figure \ref{fig:plSolMinSolMax}. 
Visible differences with the former case begin to appear only starting from the pair \#4, as shown in the lower panel of the figure: compare the green and the blue lines. 
The ratios of these lines, shown with the broken line in the lower panel, reaches $\sim 5$\% for the beam pair \#5 and $\sim 10$\% for the beam pair \#6. 
In real life, we will have only one measurement line. 
With the inevitable measurement background and uncertain contribution from the Warm Breeze, reducing the error bars and obtaining a sufficiently high fidelity in the  observations to detect deviations of the observed line from a model with the flat radial modulation function will most likely be challenging.

\section{Discussion}
\label{sec:discussion}
\noindent
The first attempt to use the direct and indirect beams to measure the ionization rate of ISN He was performed by \citet{witte_etal:96}, who used the GAS instrument on-board Ulysses \citep{witte_etal:92a}. 
The attempt was unsuccessful because the indirect beam was too weak to be clearly distinguished from the background. 
In this attempt, the equal-energy prerequisite was not postulated, which likely would have been needed because of the rather steep energy dependence of the GAS detection efficiency.
Based on the observations of the ISN He flow with IBEX-Lo, IMAP-Lo will be sufficiently sensitive to observe the indirect beam and provide multiple opportunities to compare the indirect and direct beams arriving with the same energy at the instrument.

A clear separation of the incoming ISN atoms into the direct and indirect beams is only possible within the cold-model approximation. 
In reality, with an appreciable thermal spread of the atom velocities, the atoms arrive at a given location in space from all directions, albeit with strongly non-uniform directional distribution. 
Therefore, it is unrealistic to expect that an instrument will be able to catch all of the atoms for the direct or the indirect beam. 
Consequently, direct application of Equation \ref{eq:beta0Def} for determining the ionization rate is not realistic. 
Instead, one needs to perform simulations of the exact observation conditions, similar to those presented in Section \ref{sec:implement}, and fit the ionization rates by forward modeling. 

To that end, a good grasp on the inflow parameters of the ISN gas is needed to minimize the uncertainty, as discussed in Appendix \ref{sec:betaUncert}. 
It is realistic to expect that the good knowledge of the inflow parameters that we have already now will be further improved with the future IMAP-Lo observations. 
This improvement is expected owing to the enhanced viewing geometry of this instrument, which also facilitates breaking the inflow parameter correlation, as suggested by \citet{bzowski_etal:22a} and \citet{schwadron_etal:22a}. 
With the future IMAP-Lo observations available, there is a plethora of possibilities to work on the ionization rate and inflow parameters. We can either determine the inflow parameters first and investigate the ionization rate later, or simultaneously fit the gas temperature, velocity vector, and the parameters of a model for the ionization rate based on observations collected over a long time interval during the Earth's yearly travel around the Sun while covering both the direct and indirect beams.

The contribution of the el-imp rate to the total ionization rate of ISN He is comparable to the uncertainty of the photoionization rate due to the current 15\% systematic uncertainty of the measurements of the solar EUV spectrum. 
We verified that it is feasible to expect a determination of the absolute magnitude of the ionization rate at 1 au with an accuracy exceeding the current 15\% uncertainty. 
However,  resolving the effect of the expected radial modulation of the el-imp rate will be challenging, even though the indirect atoms dive down to 0.4 au, where this modulation is expected to be strong. 

The proposed method measures the ionization rate at 1 au. 
This rate is a sum of the contributions from photoionization and el-imp ionization, with the c-x contribution negligible. 
To obtain the photoionization rate, and hence to verify the absolute calibration of the solar EUV flux, one will need an independent estimate of the el-imp rate at 1 au. 
Such an estimate can be made based on observations from another IMAP instrument, Solar Wind Electrons (SWE). 

The time resolution of the ionization rate will be approximately 0.5 y. 
This is because one can use the yearly direct and indirect observation seasons twice, i.e., combine the direct and indirect observations from a year Y, then the indirect season from Y with the direct one from year Y+1, etc. 
In principle, one can obtain the photoionization rate from just one pair of DOYs, but the need to minimize the effect of the Warm Breeze and the inevitable background suggests to use observations from a longer interval. 
The travel time inside 1 au for the atoms from the indirect beam is on the order of two months, which sets a lower limit for the time resolution from the ballistic perspective. 
The travel time of the IMAP spacecraft between the direct and indirect beam measurements suggested in Table \ref{tab:tabSimu} is much longer than that. 
On the other hand, the time interval for measurements of the indirect beam is approximately a month each year and at least in principle one can consider studying the time variation of the fitted rate during approximately one Carrington rotation period.

Existing models of the ionization rate of ISN He do show variations on a Carrington period scale. 
The travel time of ISN atoms inside 1 au is about 2 months. The travel time of IMAP (and Earth) between the direct and indirect observation sites is much longer.
Therefore, the exposures outside 1 au for the atoms observed in these two sites correspond to different epochs.
This may bias the inferred ionization rate.
It is possible, however, to mitigate this issue by performing simulations using a model where the unknown (parametrized) ionization rate is tied to variations of a proxy. 
It is well known that photoionization rate of ISN species is strongly correlated with proxies such as the F30 \citep{tanaka_kakinuma:57a, shimojo_etal:17a} or F10.7 \citep{tapping:13a} radio fluxes or to the MgIIc/w index \citep{viereck_etal:01a}. 
Hence, temporal variations of the ionization rate can be appropriately addressed in the analysis.

In principle, also the ionization rates of Ne and O could be determined using the method proposed here for ISN He, but simulations shown by \citet{sokol_etal:19c} suggest that because of their much lower abundance on one hand and their greater ionization losses on the other hand detection of their indirect beams will be very challenging on IMAP, especially during high solar activity. 
Nevertheless, a successful determination of the ionization rate for He would be insightful for other heliospheric species, at least for their photoionization rate. 
This is because H, He, Ne, O, and N are predominantly ionized by the same solar 30 nm waveband. Detecting a hypothetical bias in the photoionization rate of ISN He would suggest that appropriate corrections need to be implemented not only for He, but also for the other species. 
A similar revision would also be expected for the intensity of the solar Lyman-$\alpha$ radiation because the intensity of the solar Lyman-$\alpha$ line is strongly correlated with the 30 nm line. This revision would result in modified estimates of the solar radiation pressure for ISN H and of the intensity of the Lyman-$\alpha$ heliospheric backscatter glow \citep{kubiak_etal:21b}.

\section{Summary and conclusions}
\label{sec:conclusions}
\noindent
We have suggested a method for determination of the ionization rate of ISN He based on observations of the direct and indirect beams of ISN He performed near Earth's orbit. 
We sketched the idea of the method using the cold-model approximation. 
This idea is based on fitting the observed ratio of the direct-to-indirect beam fluxes by varying the ionization rate in the model for a precisely known temperature and inflow velocity vector of the ISN gas. 
The uncertainty will be minimized when the observation DOYs are selected to maximize the difference between the angles swept out by the direct and indirect atoms, and the impact speeds of the direct and indirect atoms are identical as seen by a neutral atom detector moving in an orbit around the Sun. 

Based on state-of-the-art simulations of the direct and indirect beams within the hot-model paradigm, using the nWTPM code, we identified the DOY pairs suitable for making these observations, and the desired elongations of the instrument boresight for these DOYs (Table \ref{tab:tabSimu}). 
We verified the pre-requisite of the equal-energy condition for the observations and identified the spin angle ranges over which the direct and indirect beams shall be integrated to mitigate the bias to the inferred ionization rate from the Warm Breeze. 
We verified that the direct/indirect flux ratio is sensitive to the ionization rate for the entire expected range of the ionization rate variation over the solar cycle. 
We also checked that the method is sensitive to the magnitude of the electron-impact ionization rate at 1 au and that it is expected to resolve the enigma of the absolute calibration of the ionization rate of ISN He, recently pointed out by \citet{swaczyna_etal:22b} based on observations from IBEX-Lo.  

The proposed method is suitable for implementation on the forthcoming IMAP mission. The expected time resolution of the ionization rate from this method is approximately 6 months. The prospective verification of the absolute magnitude of the ionization rate of ISN He is applicable also to the photoionization rates of neutral H, O, Ne, and N, which are important not only for studies of the heliosphere, but also for planetary atmospheric science. 

\appendix
\section{Appendix: Uncertainties due to the uncertainty of the inflow parameters and DOY pair selection}
\noindent
The discussion provided in this section is based on the analytic cold-model theory presented in  Section \ref{sec:baseline}.
\subsection{Uncertainty due to the uncertainty of the measured flux ratios}
\label{sec:betaUncert}
\begin{figure}
\centering
\includegraphics[width=0.45\textwidth]{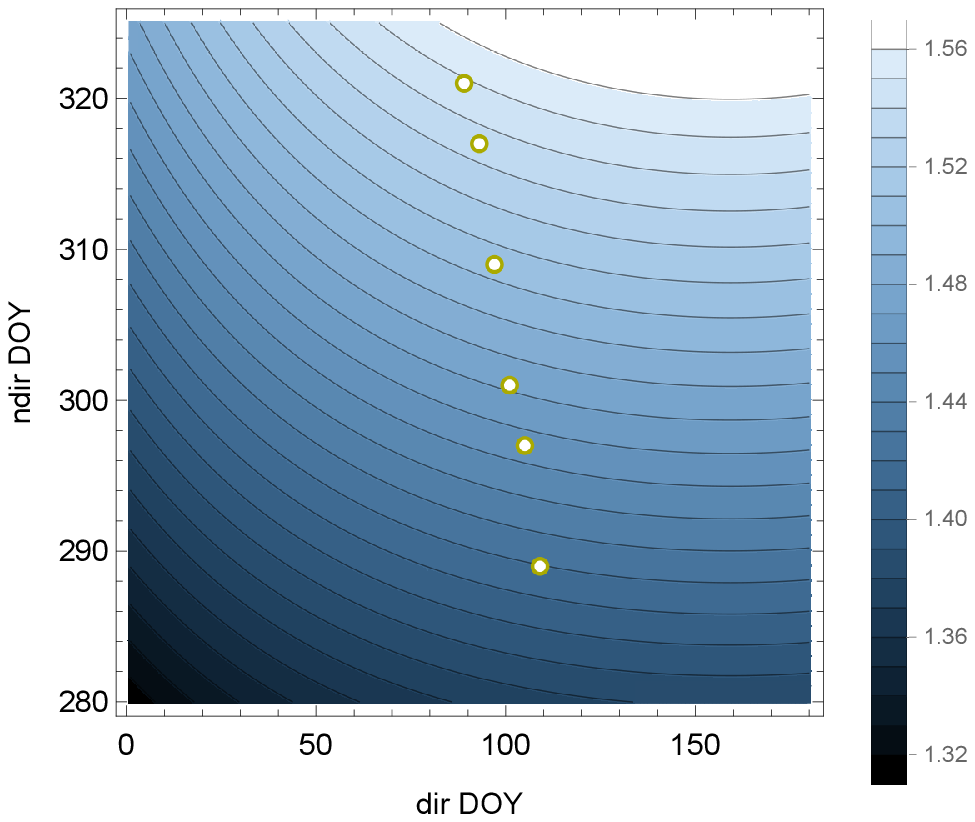}
\includegraphics[width=0.45\textwidth]{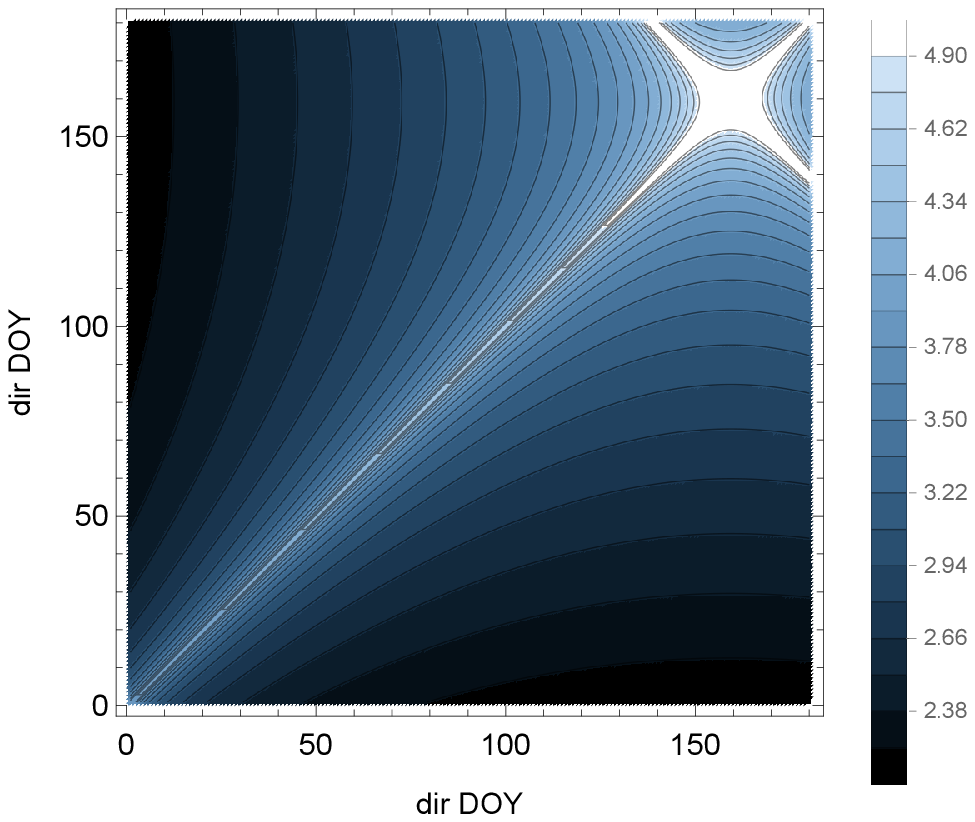}
\caption{Contribution to the relative uncertainty of the ionization rate derived from pairs of beams of ISN He: indirect and direct in the left panel, and two direct beams in the right panel. The plots present decimal logarithm of the middle term in Equation \ref{eq:fluxUncert}. The color scale is logarithmic, details are shown in the color bars. The dark color marks a low value, i.e., best suitable for determination of the ionization rate. The horizontal scale marks the DOYs for the first beam and the vertical scale those for the second beam. The yellow dots mark the selected pairs of DOYs, listed in Table \ref{tab:tabSimu}. The contributions in the right panel are symmetric relative to the equal DOYs. For identical DOYs for the direct beam, the contribution features singularity because determining the ionization rate from observations of one beam from one vantage point is impossible. See section \ref{sec:betaUncert}.}
\label{fig:pldirNdirColdUncert}
\end{figure}

\noindent
The relative uncertainty of the ionization rate $\Delta \beta_0/\beta_0$ determined from Equation \ref{eq:beta0Def} depends on the relative uncertainty $\Delta {\cal{F}}_{1,2}$ of the observed ratio ${\cal{F}}_{1,2}=F_1/F_2$ and the difference between the angles swept by the atoms on their orbits $\Delta \theta_1, \Delta \theta_2$ as follows:
\begin{equation}
  \frac{\Delta \beta_0}{\beta_0} =\frac{1}{\beta_0} \left| \frac{\partial \beta_0}{\partial {\cal{F}}_{1,2}}\right|\Delta {\cal{F}}_{1,2} =
 \frac{1}{\beta_0}\left|\frac{1}{r_E^2\left(\frac{\displaystyle \Delta \theta_2}{\displaystyle p_2} - \frac{\displaystyle \Delta \theta_1}{\displaystyle p_1} \right)}\right|  \frac{\Delta {\cal{F}}_{1,2}}{{\cal{F}}_{1,2}}.
\label{eq:fluxUncert}
\end{equation}
This implies that the uncertainty of $\beta_0$ is minimized when the expression in the denominator of Equation \ref{eq:beta0Def} is largest and the relative uncertainty of the measured flux ratio $\Delta {\cal{F}}_{1,2}/{\cal{F}}_{1,2}$ is smallest. For independent measurements of $F_1$, $F_2$,  
\begin{equation}
\Delta {\cal{F}}_{1,2}/{\cal{F}}_{1,2}=\sqrt{(\Delta F_1/F_1)^2 + (\Delta F_2/F_2)^2},
\label{eq:fluxUncert2}
\end{equation} 
so ultimately, the uncertainty of $\beta_0$ is proportional to relative uncertainties of measured fluxes $F_1, F_2$ of the two beams used.  

\subsection{Uncertainty due to the selection of pairs of vantage points}
\label{sec:fluxRatioUncert}
\noindent
Equation \ref{eq:fluxUncert}  suggests that one of the prerequisites for an accurate measurement of $\beta_0$ is minimization of the ratio $\Delta {\cal{F}}_{1,2}/{\cal{F}}_{1,2}$. 
For the determination of $\beta_0$ using the direct and indirect beams, the quantity defined in Equation \ref{eq:fluxUncert2} is dominated by the measurement uncertainty of the indirect beam. 
This beam is expected to be much lower in intensity than the direct beam and hence more challenging to measure because of the observation geometry reasons. 
Consequently, for the case of using the (direct, indirect) beams, the criterion for selection of the most suitable vantage points is based on the behavior of the middle term on the right-hand side of in Equation \ref{eq:fluxUncert}. 

We performed numerical simulations of this term for the adopted inflow parameters and the magnitudes of the ionization rate typical for ISN He. We adopted the viewing geometry similar to that used by \citet{sokol_etal:19c} to investigate the science opportunities for the IMAP-Lo experiment. 
Based on this insight, the indirect beam of ISN He can be observed between DOY 240 and $\sim 325$, while the best time for observations of the direct beam spans DOYs 1---180. 
Accordingly, we evaluated all combinations of DOYs suitable for the direct and indirect beam observations to find those promising for a minimum uncertainty. 
The results are shown in the left panel of Figure \ref{fig:pldirNdirColdUncert}. 
The results for observations of two direct beams are shown in the right panel of this figure. 

This panel suggests that if one uses two direct beams, one needs to select pairs of DOYs that are widely separated during the year. 
The smaller the DOY difference, the larger the uncertainty. 
The amplitude of the uncertainties in the figure is large, which suggests that the selection of DOY pairs should be made from the lower right or upper left quarter of the figure. 
For DOYs close to each other, the expected uncertainty of the result is large; the amplitude of the factor presented in this panel exceeds two orders of magnitude. 

The situation is different for pairs of beams composed of direct and indirect orbits, shown in the left panel of Figure \ref{fig:pldirNdirColdUncert}. 
The amplitude of the uncertainties is much lower than in the former case, only $\sim 1.7$, which suggests that practically any pair of (direct, indirect) beams is suitable. 
The best from this viewpoint are pairs from the lower-left quarter of the plot, but the advantage over selections from the other portions of this plot is small.  

Summing up this part, Figure \ref{fig:pldirNdirColdUncert} suggests that for pairs composed of indirect and direct beams, the uncertainty is typically much lower than for the pairs composed from two direct beams. 
Consequently, in the main part of the paper we focus on (indirect, direct) pairs. 
The DOY pairs selected for analysis in Section \ref{sec:implement} are presented in the left panel of Figure \ref{fig:pldirNdirColdUncert} as yellow dots.

\subsection{Uncertainty due to the uncertainty of the inflow parameters}
\label{sec:infowParUncert}
\begin{figure}
\centering
\includegraphics[width=0.45\textwidth]{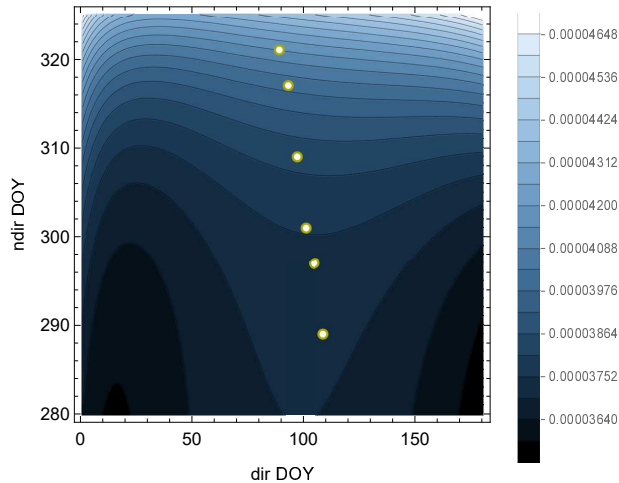}
\caption{Relative uncertainties of the ionization rate of ISN He obtained from pairs of (indirect, direct) beams listed at the vertical and horizontal axes, respectively, due to the uncertainty of the inflow parameters of ISN He. The color code is provided in the adjacent color bar. The yellow dots mark the selected pairs of DOYs, listed in Table \ref{tab:tabSimu} and used in the analysis in Section \ref{sec:implement}. Note that these uncertainties are negligibly small as compared with those due to uncertainties in the measured flux ratios, and that their amplitude is small.}
\label{fig:plflowParUncertDirDir}
\end{figure}

\noindent
Another source of the uncertainty of the derived ionization rate $\beta_0$ is the uncertainty of the inflow parameters of the ISN gas. They affect the magnitudes of all of the quantities  $\Delta \theta_{1,2}$ and $p_{1,2}$. The uncertainties of the inflow direction and speed of ISN He obtained from IBEX have been shown to form a tube in the parameter space \citep{bzowski_etal:12a, mobius_etal:12a}. \citet{schwadron_etal:22a} and \citet{bzowski_etal:22a} pointed out that this is because the observations had been collected over a relatively short arc around the Sun using an instrument that had no capability of determining the absolute speed of the impacting atoms. However, when observations are collected over a long arc or intermittently in time-separated intervals during the year, the parameter correlation can be removed. Therefore, in this analysis we assume that the uncertainties of the inflow speed and longitude are uncorrelated. The most recent analysis of the ISN He inflow direction by \citet{swaczyna_etal:22b} suggests that the speed, longitude, and latitude uncertainties are about 0.2 km~s$^{-1}$, 0.25\degr, and 0.08\degr, respectively, and we adopt these values in the assessment of the expected uncertainty in the estimates of the ionization rate.  

The relative uncertainty $\Delta \beta_0/\beta_0$ due to the uncertainties of the ISN flow vector parameters $v_B, \lambda_B$, and $\phi_B$ was calculated numerically because the analytic expressions are very complex. We used the following  formula 
\begin{equation}
\frac{ \Delta \beta_{0,x}}{\beta_0} =\frac{1}{\beta_0}\left|\frac{\partial \beta_0(x)}{\partial x}\right| \Delta x \simeq \frac{\beta_0(x + \Delta x) - \beta_0(x - \Delta x)}{2 \Delta x} \frac{\Delta x}{\beta_0},
	\label{eq:derivFormula}
\end{equation}
where $x$ is one of the inflow parameters \{$v_B, \lambda_B, \phi_B$\}, and $\beta_0$ is calculated using Equation \ref{eq:beta0Def}. Eliminating $\Delta x$ from the numerator and the denominator and using the assumption that the flow parameter uncertainties are uncorrelated, one can obtain the total relative uncertainty due to the uncertainties of the inflow parameters from the formula:
\begin{equation}
\frac{ \Delta \beta_{0}}{\beta_0} = \frac{
\displaystyle \sqrt{\sum_{x \in \{v_B, \lambda_B, \phi_B\}}\left(\beta_0(x + \Delta x) - \beta_0(x - \Delta x)\right)^2 }}
{\displaystyle 2 \beta_0}.
\label{eq:totalInflowUncert}
\end{equation}
We estimated this uncertainty based on Equation \ref{eq:totalInflowUncert} for the same combination of DOYs for the direct and indirect beams as those used for the left panel of Figure \ref{fig:pldirNdirColdUncert}. 
We adopted a magnitude of the ionization rate of ISN He equal to $10^{-7}$ s$^{-1}$, which is a typical value for the solar maximum conditions \citep{sokol_etal:20a}. 
The results are shown in Figure \ref{fig:plflowParUncertDirDir}. 
The relative uncertainty of the ionization rate due to the adopted uncertainty of the inflow parameters is very low, typically below 1\%. 
Also the variation of this uncertainty as a function of the pair selection is small. 
Given the much larger uncertainty due to the uncertainty of the measured flux ratios, it can be neglected or at least not to be considered as a weighing factor in the selection of the DOYs and ISN beams. 
The DOY pairs selected for analysis in Section \ref{sec:implement} are marked as yellow dots in the figure.

Ackonwledgments\\
The work at CBK PAN was supported by the Polish National Science Centre grant 2019/35/B/ST9/01241. The UNH contributions were supported by the NASA IBEX and IMAP projects and by NASA Grant 80NSSC18K1212.

\bibliography{ndirIonRate_v10a}{}

\begin{thebibliography}{}
\expandafter\ifx\csname natexlab\endcsname\relax\def\natexlab#1{#1}\fi
\providecommand{\url}[1]{\href{#1}{#1}}
\providecommand{\dodoi}[1]{doi:~\href{http://doi.org/#1}{\nolinkurl{#1}}}
\providecommand{\doeprint}[1]{\href{http://ascl.net/#1}{\nolinkurl{http://ascl.net/#1}}}
\providecommand{\doarXiv}[1]{\href{https://arxiv.org/abs/#1}{\nolinkurl{https://arxiv.org/abs/#1}}}

\bibitem[{{Auch{\`e}re} {et~al.}(2005){Auch{\`e}re}, {McMullin}, {Cook}, \& {et
  al.}}]{auchere_etal:05c}
{Auch{\`e}re}, F., {McMullin}, D.~R., {Cook}, J.~W., \& {et al.} 2005, in ESA
  SP-592: Solar Wind 11/SOHO 16, Connecting Sun and Heliosphere, ed. B.~Fleck,
  T.~H. Zurbuchen, \& H.~Lacoste, 327--329

\bibitem[{Axford(1972)}]{axford:72}
Axford, W.~I. 1972, in The Solar Wind, ed. J.~M.~W. C.~P.~Sonnet, P.
  J.~Coleman, NASA Spec. Publ. 308, 609--660

\bibitem[{Blum \& Fahr(1970)}]{blum_fahr:70a}
Blum, P., \& Fahr, H.~J. 1970, \aap, 4, 280

\bibitem[{{Bochsler} {et~al.}(2014){Bochsler}, {Kucharek}, {M{\"o}bius},
  {Bzowski}, {Sok{\'o}{\l}}, {Didkovsky}, \& {Wieman}}]{bochsler_etal:14a}
{Bochsler}, P., {Kucharek}, H., {M{\"o}bius}, E., {et~al.} 2014, \apjs, 210,
  12, \dodoi{10.1088/0067-0049/210/1/12}

\bibitem[{Bzowski(2008)}]{bzowski:08a}
Bzowski, M. 2008, \aap, 488, 1057, \dodoi{10.1051/0004-6361:200809393}

\bibitem[{{Bzowski} {et~al.}(2017){Bzowski}, {Kubiak}, {Czechowski}, \&
  {Grygorczuk}}]{bzowski_etal:17a}
{Bzowski}, M., {Kubiak}, M.~A., {Czechowski}, A., \& {Grygorczuk}, J. 2017,
  \apj, 845, 15.
\newblock \doarXiv{1707.02193}

\bibitem[{Bzowski {et~al.}(2022)Bzowski, Kubiak, M{\"o}bius, \&
  Schwadron}]{bzowski_etal:22a}
Bzowski, M., Kubiak, M.~A., M{\"o}bius, E., \& Schwadron, N.~A. 2022, \apj,
  938, 148, \dodoi{10.3847/1538-4357/ac8df4}

\bibitem[{Bzowski {et~al.}(2013{\natexlab{a}})Bzowski, Sok{\'o}{\l}, Kubiak, \&
  Kucharek}]{bzowski_etal:13b}
Bzowski, M., Sok{\'o}{\l}, J.~M., Kubiak, M.~A., \& Kucharek, H.
  2013{\natexlab{a}}, \aap, 557, A50, \dodoi{10.1051/0004-6361/201321700}

\bibitem[{Bzowski {et~al.}(2012)Bzowski, Kubiak, M{\"o}bius, Bochsler, Leonard,
  Heirtzler, Kucharek, Sok{\'{o}}{\l}, H{\l}ond, Crew, Schwadron, Fuselier, \&
  McComas}]{bzowski_etal:12a}
Bzowski, M., Kubiak, M.~A., M{\"o}bius, E., {et~al.} 2012, \apjs, 198, 12,
  \dodoi{10.1088/0067-0049/198/2/12}

\bibitem[{Bzowski {et~al.}(2013{\natexlab{b}})Bzowski, Sok{\'{o}}{\l},
  Tokumaru, Fujiki, Qu{\'e}merais, Lallement, Ferron, Bochsler, \&
  McComas}]{bzowski_etal:13a}
Bzowski, M., Sok{\'{o}}{\l}, J.~M., Tokumaru, M., {et~al.} 2013{\natexlab{b}},
  in {Cross-Calibration of Far {UV} Spectra of Solar Objects and the
  Heliosphere}, ed. E.~Qu{\'e}merais, M.~Snow, \& R.~Bonnet, {ISSI Scientific
  Report} No.~13 ({Springer Science+Business Media}), 67--138, doi
  10.1007/978--1--4614--6384--9$\_$3, \dodoi{10.1007/978-1-4614-6384-9_3}

\bibitem[{{Bzowski} {et~al.}(2015){Bzowski}, {Swaczyna}, {Kubiak},
  {Sok\'{o}{\l}}, {Fuselier}, {Galli}, {Heirtzler}, {Kucharek}, {Leonard},
  {McComas}, {M{\"o}bius}, {Schwadron}, \& {Wurz}}]{bzowski_etal:15a}
{Bzowski}, M., {Swaczyna}, P., {Kubiak}, M.~A., {et~al.} 2015, \apjs, 220, 28,
  \dodoi{10.1088/0067-0049/220/2/28}

\bibitem[{Bzowski {et~al.}(2019)Bzowski, Czechowski, Frisch, Fuselier, Galli,
  Grygorczuk, Heerikhuisen, Kubiak, Kucharek, McComas, M{\"o}bius, Schwadron,
  Slavin, Sok{\'o}{\l}, Swaczyna, Wurz, \& Zirnstein}]{bzowski_etal:19a}
Bzowski, M., Czechowski, A., Frisch, P., {et~al.} 2019, \apj, 882, 60,
  \dodoi{10.3847/1538-4357/ab3462}

\bibitem[{{Clette} {et~al.}(2016){Clette}, {Lef{\`e}vre}, {Cagnotti},
  {Cortesi}, \& {Bulling}}]{clette_etal:16a}
{Clette}, F., {Lef{\`e}vre}, L., {Cagnotti}, M., {Cortesi}, S., \& {Bulling},
  A. 2016, \solphys, 291, 2733, \dodoi{10.1007/s11207-016-0875-4}

\bibitem[{Fahr(1968)}]{fahr:68}
Fahr, H.~J. 1968, \apss, 2, 474

\bibitem[{Fahr(1978)}]{fahr:78}
---. 1978, \aap, 66, 103

\bibitem[{Fraternale {et~al.}(2021)Fraternale, Pogorelov, \&
  Heerikhuisen}]{fraternale_etal:21a}
Fraternale, F., Pogorelov, N.~V., \& Heerikhuisen, J. 2021, \apjl, 921, L24,
  \dodoi{10.3847/2041-8213/ac313c}

\bibitem[{{Galli} {et~al.}(2015){Galli}, {Wurz}, {Park}, {Kucharek},
  {M{\"o}bius}, {Schwadron}, {Sok\'{o}{\l}}, {Bzowski}, {Kubiak}, {Swaczyna},
  {Fuselier}, \& {McComas}}]{galli_etal:15a}
{Galli}, A., {Wurz}, P., {Park}, J., {et~al.} 2015, \apjs, 220, 30,
  \dodoi{10.1088/0067-0049/220/2/30}

\bibitem[{Holzer(1977)}]{holzer:77}
Holzer, T.~E. 1977, Rev. Geophys., 15, 467, \dodoi{10.1029/RG015i004p00467}

\bibitem[{{King} \& {Papitashvili}(2005)}]{king_papitashvili:05}
{King}, J.~H., \& {Papitashvili}, N.~E. 2005, \jgr, 110, 2104,
  \dodoi{10.1029/2004JA010649}

\bibitem[{Kubiak {et~al.}(2021)Kubiak, Bzowski, Kowalska-Leszczynska, \&
  Strumik}]{kubiak_etal:21b}
Kubiak, M.~A., Bzowski, M., Kowalska-Leszczynska, I., \& Strumik, M. 2021,
  \apjs, 254, 17, \dodoi{10.3847/1538-4365/abeb78}

\bibitem[{{Kubiak} {et~al.}(2014){Kubiak}, {Bzowski}, {Sok{\'o}{\l}},
  {Swaczyna}, {Grzedzielski}, {Alexashov}, {Izmodenov}, {Moebius}, {Leonard},
  {Fuselier}, {Wurz}, \& {McComas}}]{kubiak_etal:14a}
{Kubiak}, M.~A., {Bzowski}, M., {Sok{\'o}{\l}}, J.~M., {et~al.} 2014, \apjs,
  213, 29, \dodoi{10.1088/0067-0049/212/2/29}

\bibitem[{Kubiak {et~al.}(2016)Kubiak, Swaczyna, Bzowski, Sok{\'o}{\l},
  Fuselier, Galli, Heirtzler, Kucharek, Leonard, McComas, Park, Schwadron, \&
  Wurz}]{kubiak_etal:16a}
Kubiak, M.~A., Swaczyna, P., Bzowski, M., {et~al.} 2016, \apjs, 223, 35,
  \dodoi{10.1088/0067-0049/220/2/35}

\bibitem[{Lee {et~al.}(2012)Lee, Kucharek, M{\"o}bius, Wu, Bzowski, \&
  McComas}]{lee_etal:12a}
Lee, M.~A., Kucharek, H., M{\"o}bius, E., {et~al.} 2012, \apjs, 198, 10,
  \dodoi{10.1088/0067-0049/198/2/10}

\bibitem[{Machol {et~al.}(2019)Machol, Snow, Woodraska, Woods, Viereck, \&
  Coddington}]{machol_etal:19a}
Machol, J.~L., Snow, M., Woodraska, D., {et~al.} 2019, Earth and Space Science,
  6, 2263, \dodoi{10.1029/2019EA000648}

\bibitem[{McComas {et~al.}(2018)McComas, Christian, Schwadron, Fox, Westlake,
  Allegrini, Baker, Biesecker, Bzowski, Clark, Cohen, Cohen, Dayeh, Decker,
  de~Nolfo, Desai, andH.A. Elliott, Fahr, Frisch, Funsten, Fuselier, Galli,
  Galvin, Giacalone, Gkioulidou, Guo, Horanyi, Isenberg, Janzen, Kistler,
  Korreck, Kubiak, Kucharek, Larsen, Leske, Lugaz, Luhmann, Matthaeus, Mitchel,
  Moebius, Ogasawara, Reisenfeld, Richardson, Russell, Sok{\'o}{\l}, Spence,
  Skoug, Sternovsky, Swaczyna, Szalay, Tokumaru, andP. Wurz, Zank, \&
  Zirnstein}]{mccomas_etal:18b}
McComas, D., Christian, E., Schwadron, N., {et~al.} 2018, \ssr, 214, 116,
  \dodoi{10.1007/s11214-018-0550-1}

\bibitem[{{McComas} {et~al.}(2009){McComas}, {Allegrini}, {Bochsler},
  {Bzowski}, {Collier}, {Fahr}, {Fichtner}, {Frisch}, {Funsten}, {Fuselier},
  {Gloeckler}, {Gruntman}, {Izmodenov}, {Knappenberger}, {Lee}, {Livi},
  {Mitchell}, {M{\"o}bius}, {Moore}, {Pope}, {Reisenfeld}, {Roelof},
  {Scherrer}, {Schwadron}, {Tyler}, {Wieser}, {Witte}, {Wurz}, \&
  {Zank}}]{mccomas_etal:09a}
{McComas}, D.~J., {Allegrini}, F., {Bochsler}, P., {et~al.} 2009, \ssr, 146,
  11, \dodoi{10.1007/s11214-009-9499-4}

\bibitem[{{M{\"o}bius} {et~al.}(2004){M{\"o}bius}, {Bzowski}, {Chalov}, {Fahr},
  {Gloeckler}, {Izmodenov}, {Kallenbach}, {Lallement}, {McMullin}, {Noda},
  {Oka}, {Pauluhn}, {Raymond}, {Ruci{\'n}ski}, {Skoug}, {Terasawa}, {Thompson},
  {Vallerga}, {von Steiger}, \& {Witte}}]{mobius_etal:04a}
{M{\"o}bius}, E., {Bzowski}, M., {Chalov}, S., {et~al.} 2004, \aap, 426, 897,
  \dodoi{10.1051/0004-6361:20035834}

\bibitem[{{M{\"o}bius} {et~al.}(2012){M{\"o}bius}, {Bochsler}, {Bzowski},
  {Heirtzler}, {Kubiak}, {Kucharek}, {Lee}, {Leonard}, {Schwadron}, {Wu},
  {Fuselier}, {Crew}, {McComas}, {Petersen}, {Saul}, {Valovcin}, {Vanderspek},
  \& {Wurz}}]{mobius_etal:12a}
{M{\"o}bius}, E., {Bochsler}, P., {Bzowski}, M., {et~al.} 2012, \apjs, 198, 11,
  \dodoi{10.1088/0067-0049/198/2/11}

\bibitem[{Porowski {et~al.}(2022)Porowski, Bzowski, \&
  Tokumaru}]{porowski_etal:22a}
Porowski, C., Bzowski, M., \& Tokumaru, M. 2022, \apjs, 259, 2,
  \dodoi{10.3847/1538-4365/ac35d7}

\bibitem[{Ruci{\'n}ski {et~al.}(2003)Ruci{\'n}ski, Bzowski, \&
  Fahr}]{rucinski_etal:03}
Ruci{\'n}ski, D., Bzowski, M., \& Fahr, H.~J. 2003, \ag, 21, 1315,
  \dodoi{10.5194/angeo-21-1315-2003}

\bibitem[{Ruci{\'n}ski {et~al.}(1996)Ruci{\'n}ski, Cummings, Gloeckler,
  Lazarus, M{\"o}bius, \& Witte}]{rucinski_etal:96a}
Ruci{\'n}ski, D., Cummings, A.~C., Gloeckler, G., {et~al.} 1996, \ssr, 78, 73,
  \dodoi{10.1007/BF00170794}

\bibitem[{Ruci{\'n}ski \& Fahr(1989)}]{rucinski_fahr:89}
Ruci{\'n}ski, D., \& Fahr, H.~J. 1989, \aap, 224, 290

\bibitem[{Ruci{\'n}ski \& Fahr(1991)}]{rucinski_fahr:91}
---. 1991, \ag, 9, 102

\bibitem[{Schwadron {et~al.}(2022)Schwadron, M{\"o}bius, McComas, Bower, Bower,
  Bzowski, Fuselier, Heirtzler, Kubiak, Lee, Rahmanifard, Sok{\'o}{\l},
  Swaczyn, \& Winslow}]{schwadron_etal:22a}
Schwadron, N.~A., M{\"o}bius, E., McComas, D.~J., {et~al.} 2022, \apjs, 258, 7,
  \dodoi{10.3847/1538-4365/ac2fa9}

\bibitem[{{Shimojo} {et~al.}(2017){Shimojo}, {Iwai}, {Asai}, {Nozawa},
  {Minamidani}, \& {Saito}}]{shimojo_etal:17a}
{Shimojo}, M., {Iwai}, K., {Asai}, A., {et~al.} 2017, \apj, 848, 62,
  \dodoi{10.3847/1538-4357/aa8c75}

\bibitem[{Snow {et~al.}(2019)Snow, Machol, Viereck, Woods, Weber, Woodraska, \&
  Elliott}]{snow_etal:19a}
Snow, M., Machol, J., Viereck, R., {et~al.} 2019, Earth and Space Science, 2,
  \dodoi{10.1029/2019EA000652}

\bibitem[{{Snow} {et~al.}(2013){Snow}, {Reberac}, {Qu{\'e}merais}, {Clarke},
  {McClintock}, \& {Woods}}]{snow_etal:13}
{Snow}, M., {Reberac}, A., {Qu{\'e}merais}, E., {et~al.} 2013, in ISSI
  Scientific Report Series, Vol.~13, Cross-Calibration of Far UV Spectra of
  Solar System Objects and the Heliosphere, ed. E.~{Qu{\'e}merais}, M.~{Snow},
  \& R.-M. {Bonnet} ({Springer Science+Business Media}), 191--226,
  \dodoi{10.1007/978-1-4614-6384-9_7}

\bibitem[{{Sok{\'o}{\l}} {et~al.}(2019{\natexlab{a}}){Sok{\'o}{\l}}, {Bzowski},
  \& {Tokumaru}}]{sokol_etal:19a}
{Sok{\'o}{\l}}, J.~M., {Bzowski}, M., \& {Tokumaru}, M. 2019{\natexlab{a}},
  \apj, 872, 57, \dodoi{10.3847/1538-4357/aaf737}

\bibitem[{{Sok{\'o}{\l}} {et~al.}(2019{\natexlab{b}}){Sok{\'o}{\l}}, {Kubiak},
  {Bzowski}, {M{\"o}bius}, \& {Schwadron}}]{sokol_etal:19c}
{Sok{\'o}{\l}}, J.~M., {Kubiak}, M.~A., {Bzowski}, M., {M{\"o}bius}, E., \&
  {Schwadron}, N. 2019{\natexlab{b}}, \apjs, 245, 28,
  \dodoi{10.3847/1538-4365/ab50bc}

\bibitem[{{Sok\'{o}{\l}} {et~al.}(2015{\natexlab{a}}){Sok\'{o}{\l}}, {Kubiak},
  {Bzowski}, \& {Swaczyna}}]{sokol_etal:15b}
{Sok\'{o}{\l}}, J.~M., {Kubiak}, M.~A., {Bzowski}, M., \& {Swaczyna}, P.
  2015{\natexlab{a}}, \apjs, 220, 27, \dodoi{10.1088/0067-0049/220/2/27}

\bibitem[{{Sok{\'o}{\l}} {et~al.}(2020){Sok{\'o}{\l}}, {McComas}, {Bzowski}, \&
  {Tokumaru}}]{sokol_etal:20a}
{Sok{\'o}{\l}}, J.~M., {McComas}, D.~J., {Bzowski}, M., \& {Tokumaru}, M. 2020,
  \apj, 897, 179, \dodoi{10.3847/1538-4357/ab99a4}

\bibitem[{{Sok\'{o}{\l}} {et~al.}(2015{\natexlab{b}}){Sok\'{o}{\l}}, {Bzowski},
  {Kubiak}, {Swaczyna}, {Galli}, {Wurz}, {M{\"o}bius}, {Kucharek}, {Fuselier},
  \& {McComas}}]{sokol_etal:15a}
{Sok\'{o}{\l}}, J.~M., {Bzowski}, M., {Kubiak}, M.~A., {et~al.}
  2015{\natexlab{b}}, \apjs, 220, 29, \dodoi{10.1088/0067-0049/220/2/29}

\bibitem[{{Strumik} {et~al.}(2021){Strumik}, {Bzowski}, \&
  {Kubiak}}]{strumik_etal:21b}
{Strumik}, M., {Bzowski}, M., \& {Kubiak}, M.~A. 2021, \apjl, 919, L18,
  \dodoi{10.3847/2041-8213/ac2734}

\bibitem[{{Svalgaard} \& {Hudson}(2010)}]{svalgaard_hudson:10a}
{Svalgaard}, L., \& {Hudson}, H.~S. 2010, in Astronomical Society of the
  Pacific Conference Series, Vol. 428, SOHO-23: Understanding a Peculiar Solar
  Minimum, ed. {S.~R.~Cranmer, J.~T.~Hoeksema, \& J.~L.~Kohl}, 325.
\newblock \doarXiv{1003.4281}

\bibitem[{{Swaczyna} {et~al.}(2022){Swaczyna}, {Eddy}, {Zirnstein}, {Dayeh},
  {McComas}, {Funsten}, \& {Schwadron}}]{swaczyna_etal:22a}
{Swaczyna}, P., {Eddy}, T.~J., {Zirnstein}, E.~J., {et~al.} 2022, \apjs, 258,
  6, \dodoi{10.3847/1538-4365/ac2f47}

\bibitem[{Swaczyna {et~al.}(2018)Swaczyna, Bzowski, Kubiak, Sok{\'o}{\l},
  Fuselier, Galli, Heirtzler, Kucharek, McComas, M{\"o}bius, Schwadron, \&
  Wurz}]{swaczyna_etal:18a}
Swaczyna, P., Bzowski, M., Kubiak, M.~A., {et~al.} 2018, \apj, 854, 119,
  \dodoi{10.3847/1538-4357/aaabbf}

\bibitem[{Swaczyna {et~al.}(2022)Swaczyna, Kubiak, Bzowski, Bower, Fuselier,
  Galli, Heirtzler, McComas, M{\"o}bius, Rahmanifard, \&
  Schwadron}]{swaczyna_etal:22b}
Swaczyna, P., Kubiak, M.~A., Bzowski, M., {et~al.} 2022, \apjs, 259, 42,
  \dodoi{10.3847/1538-4365/ac4bde}

\bibitem[{Tanaka \& Kakinuma(1957)}]{tanaka_kakinuma:57a}
Tanaka, H., \& Kakinuma, T. 1957, Proceedings of the Research Institute of
  Atmospherics, 4, 60

\bibitem[{{Tapping} \& {Vald{\'e}s}(2011)}]{tapping_valdes:11a}
{Tapping}, K., \& {Vald{\'e}s}, J. 2011, \solphys, 272, 337,
  \dodoi{10.1007/s11207-011-9827-1}

\bibitem[{{Tapping}(2013)}]{tapping:13a}
{Tapping}, K.~F. 2013, Space Weather, 11, 1, \dodoi{10.1002/swe.20064}

\bibitem[{Thomas(1978)}]{thomas:78}
Thomas, G.~E. 1978, Ann. Rev. Earth Planet. Sci., 6, 173,
  \dodoi{10.1146/annurev.ea.06.050178.001133}

\bibitem[{Tokumaru {et~al.}(2021)Tokumaru, Fujiki, Kojima, \&
  Iwai}]{tokumaru_etal:21a}
Tokumaru, M., Fujiki, K., Kojima, M., \& Iwai, K. 2021, \apj, 922, 73,
  \dodoi{10.3847/1538-4357/ac1862}

\bibitem[{Viereck {et~al.}(2001)Viereck, Puga, McMullin, \&
  Tobiska}]{viereck_etal:01a}
Viereck, R., Puga, L., McMullin, D., \& Tobiska, W. 2001, \grl, 28, 1343,
  \dodoi{10.1029/2000GL012551}

\bibitem[{{Wieman} {et~al.}(2014){Wieman}, {Didkovsky}, \&
  {Judge}}]{wieman_etal:14a}
{Wieman}, S.~R., {Didkovsky}, L.~V., \& {Judge}, D.~L. 2014, \solphys, 289,
  2907, \dodoi{10.1007/s11207-014-0519-5}

\bibitem[{Witte {et~al.}(1996)Witte, Banaszkiewicz, \&
  Rosenbauer}]{witte_etal:96}
Witte, M., Banaszkiewicz, M., \& Rosenbauer, H. 1996, \ssr, 78, 289,
  \dodoi{10.1007/BF00170815}

\bibitem[{{Witte} {et~al.}(1992){Witte}, {Rosenbauer}, {Keppler}, {Fahr},
  {Hemmerich}, {Lauche}, {Loidl}, \& {Zwick}}]{witte_etal:92a}
{Witte}, M., {Rosenbauer}, H., {Keppler}, E., {et~al.} 1992, \aaps, 92, 333

\bibitem[{{Woods} \& {DeLand}(2021)}]{woods_deland:21a}
{Woods}, T.~N., \& {DeLand}, M.~T. 2021, Earth and Space Science, 8, e01740,
  \dodoi{10.1029/2021EA001740}

\bibitem[{{Woods} {et~al.}(2005){Woods}, {Eparvier}, {Bailey}, {Chamberlin},
  {Lean}, {Rottman}, {Solomon}, {Tobiska}, \& {Woodraska}}]{woods_etal:05a}
{Woods}, T.~N., {Eparvier}, F.~G., {Bailey}, S.~M., {et~al.} 2005, \jgr, 110,
  A01312, \dodoi{10.1029/2004JA010765}

\end{thebibliography}
\bibliographystyle{aasjournal}  

\end{document}